\documentclass[showpacs,preprintnumbers,amsmath,amssymb,floatfix]{revtex4-1}

\usepackage{color}   
\usepackage{graphicx}
\usepackage{dcolumn}
\usepackage{bm}

\usepackage[update,prepend]{epstopdf} 

\begin{document}

\title{Stability of Matter-Antimatter Molecules} 

\author{Cheuk-Yin Wong${}^*$\footnote[0]{${}^*$Corresponding Author\\ Email address: ${}^*$wongc@ornl.gov,${}^a$teck@physics.auburn.edu }} 

\affiliation{Physics Division, Oak Ridge National Laboratory, Oak Ridge, 
TN 37831}

\author{ Teck-Ghee Lee${}^a$}

\affiliation{Department of Physics,  Auburn University, Auburn, AL 36849}

\def\bbox#1{\hbox{\boldmath${#1}$}}
\def\bb    #1{\hbox{\boldmath${#1}$}}
\def\blambda{{\hbox{\boldmath $\lambda$}}}
\def\eeta{{\hbox{\boldmath $\eta$}}}
\def\bxi{{\hbox{\boldmath $\xi$}}}

\begin{abstract}

We examine the stability of matter-antimatter molecules by reducing
the four-body problem into a simpler two-body problem with residual
interactions. We find that matter-antimatter molecules with
constituents $(m_1^+,m_2^-,{\bar m}_2^+,{\bar m_1}^-)$ possess bound
states if their constituent mass ratio $m_1/m_2$ is greater than
about 4.  This stability condition suggests that the binding of
matter-antimatter molecules is a rather common phenomenon.  We
evaluate the binding energies and eigenstates of matter-antimatter
molecules $(\mu^+e^-)$-$(e^+\mu^-)$, $(\pi^+e^-)$-$(e^+\pi^-)$,
$(K^+e^-)$-$(e^+K^-)$, $(pe^-)$-$(e^+\bar p)$, $(p\mu^-)$-$(\mu^+{\bar
  p})$, and $(K^+\mu^-)$-$(\mu^+K^-)$, which satisfy the stability
condition.  We estimate the molecular annihilation lifetimes in their
$s$ states.

\end{abstract}


\maketitle

\section{Introduction}

The study of the stability and the properties of matter-antimatter
molecules has a long history, starting with the pioneering work of
Wheeler who suggested in 1946 that $(e^+e^-)$ might be bound with its
antimatter partner to form a matter-antimatter molecule \cite{Whe46}.
Since then, the problem has been examined theoretically by many
workers
\cite{Hyl47,Mor73,Ho83,Gri89,Nus93,Koz93,Zyg04,Kolos,Armour,Froelich,Jon01,Strasburger,Labzowsky,Sharipov}
(for a review and other references see \cite{ Cha01}).  Wheeler went
on to explore the properties of an assembly of $(e^+e^-)^n$ atoms and
molecules if they were made, and he outlined the phase boundaries in
temperature and pressure separating various phases of $(e^+e^-)$ atoms
and $(e^+e^-)$-$(e^-e^+)$ molecules in their gaseous, liquid,
super-fluid, crystal, and metallic states \cite{Whe88}.  However, the
experimental detection of matter-antimatter molecules is difficult,
and the $(e^+e^-)$-$(e^-e^+)$ molecule was successfully detected only
recently, as late as 2007 \cite{Cas07}.

In spite of extensive past investigations, our knowledge of
matter-antimatter molecules remains rather incomplete, being limited
to $(e^+)^m(e^-)^n$ and some aspects of $(p e^-)$-$(e^+{\bar p})$.
There are however stable and meta-stable charged particles and
antiparticles, such as $e^\pm$, $p\bar p$, $\mu^\pm$, $\pi^\pm$,
$K^\pm $, $\tau^\pm$, etc.  The conditions for the molecular binding
of four-body particle-antiparticle complexes containing these charged
constituents are not known, nor are their annihilation lifetimes, if
these matter-antimatter molecules turn out to be bound.

With the advent of the Relativistic Heavy-Ion Collider at Brookhaven
and the Large hadron Collider at CERN, a large number of charged
particles and antiparticles are produced in high-energy $pp$ and
heavy-ion collisions.  The production of matter and antimatter
particles in close space-time proximity raises the interesting
question whether chance encounters of some of the produced charged
particles and antiparticles may lead to the formation of
matter-antimatter molecules as debris of the collision.  The detection
of matter-antimatter molecules in high-energy nuclear collisions will
need to overcome the difficulty of large combinatorial background
noises that may be present.  In another related area, recent
production and trapping of cold antihydrogen
\cite{Amoretti,Gabrielse,And10} provide the possibility of bringing
matter and antimatter atoms close together.  Furthermore,
electron-positron colliders with fine energy resolutions may be used
to produce stable matter-antimatter molecules as resonances with
finite widths, when the colliding $e^+$ and $e^-$ combination has the
same quantum number as the matter-antimatter molecules.

The detection of new matter-antimatter molecules is however a
difficult task, as evidenced by the long span of time between the
proposal and the observation of the $(e^+e^-)$-$(e^-e^+)$ molecule.
Additional instrumentation and experimental apparatus may be needed.
It is nonetheless an interesting theoretical question to investigate
systematically the general factors affecting the stability of
matter-antimatter molecules, whether any of these matter-antimatter
molecules may be bound, and if they are bound, what are their binding
energies, annihilation lifetimes, and other characteristics. Answers
to these questions will help us assess whether it may ever be feasible
to detect them experimentally in the future.

Following Wheeler \cite{Whe46}, we shall use the term ``atom" to
represent a two-body bound state of a positive and a negative charged
pair that can form a building block, out of which more complex
``molecules" can be constructed.  In the present work, we shall limit
our attention to molecules in which the four constituents $\{m_1$,
$m_2$, $m_3$, $m_4\}$ consist of two charge conjugate pairs, with
$m_3$ the charge conjugate of $m_2$, and $m_4$ the charge conjugate of
$m_1$.  We shall arrange and order the constituents according to their
masses such that $m_1$$>$$m_2$, and the charges of $m_1$ and $m_2$ be
$+e$ and $-e$ respectively.  To make the problem simple, we shall
consider molecules containing non-identical constituents $m_1$ and
$m_2$ and their antiparticle counterparts, such that there are no
identical particles among the constituents.  Molecules with identical
constituents require additional considerations on the symmetries of
the wave function with respect to the exchange of the pair of
identical particles, which are beyond the scope of the present
investigation.

To study the structure of the molecules, we shall consider only
Coulomb interactions between particles and neglect strong
interactions, as the range of strong interactions is considerably
smaller than the Bohr radius of the relevant particle-antiparticle
system.

Previously, based on the method proposed for the study of molecular
states in heavy quark mesons \cite{Won04}, we obtained the interatomic
potential for the $H\bar H$ system \cite{Lee08}.  We shall generalize
our consideration to cases of constituent particles of various masses
and types and shall quantize the four-body Hamiltonian to obtain
molecular eigenstates of the four-body system.  If molecular states
are found, we shall determine their annihilation lifetimes and their
spin dependencies, if any.

It is worth pointing out that the subject matter of molecular states
appears not only in atomic and molecular physics, but also in nuclear
physics and hadron spectroscopy.  Wheeler's 1937 article entitled
``Molecular Viewpoints in Nuclear Structure" introduced molecular
physics concepts such as resonating groups and alpha particle groups
to nuclear physics \cite{Whe37}.  Indeed, nucleus-nucleus molecular
states have been observed previously in the collision of light nuclei
near the Coulomb barrier \cite{Alm60}.  Molecular states of
heavy-quark mesons have been proposed in high-energy hadron
spectroscopy \cite{Won04,Tor03,Clo04,Bra04,Swa04} to explain the
narrow 3872 MeV state discovered by the Belle Collaboration
\cite{Cho03} and other Collaborations \cite{Aco04}.  The general
stability condition established here for the Coulomb four-body problem
for matter-antimatter molecules may have interesting implications or
generalizations in other branches of physics.

This paper is organized as follows. In Section II, we review the
families of states of the four-body matter-antimatter system so as to
introduce the method of our investigation.  In Section III, the
mathematical details of our formulation are presented and the
four-body problem is reduced to a simple two-body problem in terms of
the interaction of two atoms with residual interactions.  The
interaction potential is found to consist of the sum of the direct
potential $V_{\rm dir}$ and the polarization potential $V_{\rm pol}$.
In Section IV, we show how to evaluate the interaction matrix
elements.  In Section V, we show the analytical result for the direct
potential $V_{\rm dir}$.  In Section VI, the polarization potential is
evaluated for the virtual excitation to the complete set of bound and
continuum atomic states.  The results of the interaction potential for
$(pe^+)$-$(e^+{\bar p})$ are discussed in Section VII.  The
interaction potentials for other molecular systems are examined in
Section VIII.  In Section IX, we solve the Schr\" odinger equation for
molecular motion and obtain the molecular eigenstates for different
systems.  We discuss the annihilation rates and lifetimes of the
molecular states in Section X.  Finally, Section XI gives some
discussions and conclusions of the present work. Some of the details
of the analytical results are presented in the Appendix.

\section{Families of four-particle states}

The four constituent particles can be arranged in different ways
leading to different types of states.  There is the $A{\bar A}$ family
of states of the type $A(m_1^+ m_2^-)$-${\bar A} ({\bar m}_2^+ {\bar
  m_1}^-)$, in which $m_1^+$ and $m_2^-$ orbit around each other to
form the $A_\nu (m_1^+ m_2^-)$ atom in the $\nu$ state, while ${\bar
  m}_2^+$ and ${\bar m}_1^-$ orbit around each other to form the
${\bar A}_{\nu'} ({\bar m}_2^+ {\bar m}_1^-)$ atom in the ${\nu'}$
state.  When $A_\nu$ and ${\bar A}_{\nu'}$ are separated at large
distances, their asymptotic state energy is
\begin{eqnarray}
E_{A\bar A}(n_\nu,n_{\nu'})=-\alpha^2 \frac{m_1 m_2}{ m_1+m_2}
\left (\frac{1}{n_\nu^2} +\frac{1}{n_{\nu'}^2}\right ) .
\end{eqnarray}
There is another $M\bar M$ family of states of the type $M (m_1^+
{\bar m_1}^-)$-${\bar M}({\bar m_2}^+ m_2^-)$, in which $m_1^+$ and
${\bar m}_1^-$ orbit around each other to form the $M_\nu (m_1^+ {\bar
  m}_1^-)$ atom in the $\nu$ state, while ${\bar m}_2^+$ and $m_2^-$
orbit around each other to form the ${\bar M}_{\nu'} ({\bar m}_2^+
m_2^-)$ atom in the $\nu'$ state.  Their asymptotic state energy is
\begin{eqnarray}
E_{M{\bar M}}(n_\nu,n_{\nu'})=- \frac{1}{4} \alpha^2\left ( \frac{m_1}{n_\nu^2} +\frac{m_2}{n_{\nu'}^2}\right ).
\end{eqnarray} 
 For the same values of $n_\nu=n_{\nu'}$, the asymptotic state of the
 $M{\bar M}$ family lies lower in energy than the asymptotic state of
 the $A\bar A$ family, except when $m_1=m_2$ for which they are at the
 same level,
\begin{eqnarray}
E_{M{\bar M}}(n_\nu,n_{\nu}) < E_{A\bar A}  (n_\nu,n_{\nu}) &{\rm~~if~~}&  m_1\ne m_2\nonumber\\
{\rm and~~~}E_{M{\bar M}}(n_\nu,n_{\nu}) = E_{A\bar A}  (n_\nu,n_{\nu}) &{\rm~~if~~}&  m_1= m_2.
\end{eqnarray}
On the other hand, by varying the principal quantum numbers, many of
the asymptotic states of one family can lie close to the energy levels
of the asymptotic states of the other family.  Level crossing between
states and the mixing of states of different families can occur when
the atoms are brought in close proximity to each other.

There are two different methods to study molecular states.  In the
first method, one reduces the four-body problem into a simpler
two-body problem.  One breaks up the four-body Hamiltonian into the
unperturbed Hamiltonians of two atoms, plus residual interactions and
the kinetic energies of the atoms.  The unperturbed Hamiltonians of
the two atoms can be solved exactly.  Using the atomic states as
separable two-body basis, one constructs molecular states with the
atoms as simple building blocks and quantize the four-particle
Hamiltonian \cite{Won04}.  The quantized eigenstate obtained in such a
method may not necessarily contain all the correlations.  They may
also not necessarily be the lowest states of the four-body system.
They however have the advantage that the center-of-mass motion of the
composite atoms are properly treated and the formulation can be
applied to systems with vastly different mass ratios $m_1/m_2$.  They
provide a clear and simple picture of the molecular structure.  They
also provide vital information on the condition of molecular stability
and the values of molecular eigenenergies.  The knowledge of the
molecular eigenfunctions provides information on other properties of
the molecular states and their annihilation lifetimes.  Furthermore,
these molecular states can form doorways for states of greater
complexity with additional correlations.  For example, one can
multiply the four-body wave functions of a molecular state obtained in
such a method (see Eq.\ (\ref{wf}) below) by a complete set of
correlated wave functions for a particular pair of constituents, and
diagonalize the four-body Hamiltonian with such a basis.  The
eigenstates obtained after such a diagonalization represent the
splitting of the doorway state into finer molecular states containing
additional correlations.  As the method exhibits a clear molecular
structure in terms of composite two-body objects, its asymptotic
states illustrate how the molecule may be formed by the collision of
the composite atoms.  Finally, if one depicts the orbiting of one
particle relative to another particle as a ``dance pattern'' with the
topology of a ring, then the dance patterns of the different
constituents and atoms in different families have distinctly different
connectivities and topological structures.  The transition of a state
from one family to another family will involve the breaking of one
type of dance pattern and re-establishing another type of dance
pattern.  It is reasonable to conjecture that their distinct
topological structures may suppress the transition amplitude between
families and may allow the atoms to retain some of their
characteristics and stability in their dynamical motion and
transitions.

There is an alternative second method to study the molecular states by
taking the interaction potential to be the adiabatic potential
obtained in a variational calculation for the lowest-energy state of
the four particle complex in the Born-Oppenheimer potential, in which
the positions of two heavy constituents are held fixed
\cite{Zyg04,Kolos,Armour,Froelich,Jon01,Strasburger,Labzowsky,Sharipov}.
It should be realized that such variational calculations have not yet
been fully variational.  By fixing the positions of the heavy
constituents (the proton and antiproton in the case of the $(p e^- e^+
{\bar p})$ complex), the trial wave functions of the heavy
constituents have been constrained to be delta functions without
variations.  If the motion of the heavy constituents were allowed to
vary in a fully variational calculation, the lowest energy state would
be the one in which the heavy constituents would orbit around each
other in their atomic orbitals, with binding energies proportional to
their heavy masses.  The variational calculations of the higher
molecular states will need to insure the orthogonality of the state
relative to lower lying ones.  Furthermore, as the variational
calculation searches for the state with the lowest energy, motion in
the relative coordinates between the heavier masses corresponds to
constraining a trajectory along a path with adiabatic transitions
whenever a level crossing occurs.  In a collective molecular motion,
these level crossing may not necessarily be adiabatic \cite{Hil53}
because the speed of the collective motion becomes large at small
interatomic separations, and the transition matrix element between
different families may be suppressed due to the difference in their
topological structures.  It is reasonable to suggest that while
variational calculations may provide useful information on the
four-particle complex, they should not be the only method to examine
the molecular structure of the four-body system.

We shall use the first method to study the eigenstates of the
four-body system and construct states built on the $A\bar A$ family.
Accordingly, we break the four-body Hamiltonian into a part containing
the unperturbed Hamiltonians of the $A$ and ${\bar A}$ atoms and
another part containing the residual interactions and the kinetic
energies of $A$ and $\bar A$. The quantization of the four-body
Hamiltonian then gives the molecular states of interest, as in a
previous study of molecular states in heavy mesons in hadron physics
\cite{Won04}. Such a separation of the four-body Hamiltonian is
justified because the composite $A$ and $\bar A$ atoms are neutral
objects, and their residual interaction $V_I$ between the constituents
of $A$ and $\bar A$ involve many cancellations.  As a consequence, the
non-diagonal matrix element for excitation arising from the residual
interaction is relatively small in comparison with the difference
between unperturbed state energies, and the perturbation expansion is
expected to converge.

Molecular states can be constructed using composite objects of $A_\nu$
and ${\bar A}_{\nu'}$ in various asymptotic $\nu$ and $\nu'$ states.
We shall be interested in molecular states in which the building block
atoms $A_0$ and ${\bar A}_0$ are in their ground states at asymptotic
separations.  As the two ground state atoms approach each other, they
will be excited and polarized and their virtual excitation will lead
to an interaction potential between the atoms, from which the
eigenstates of the molecule will be determined.  

Molecular states constructed with excited $A_\nu $ and ${\bar
  A}_{\nu'}$ atoms can be similarly considered in a simple
generalization in the future.  Such a possibility brings into focus
the richness of states in the four-body system, as the molecular
states will have different interatomic interactions, obey different
stability conditions, and have different properties with regard to
annihilation and production.

\section{The Four-body Problem in a Separable Two-Body Basis}

\begin{figure} [h]
\hspace*{0.0cm}
\includegraphics[scale=0.45]{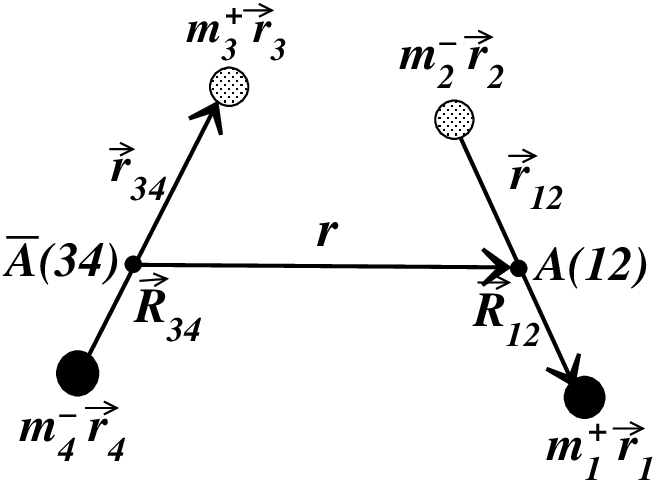}
\caption{ The coordinates of the $\{m_1^+,m_2^-,m_3^+,m_4^- \}$
system.  } \label{fig1}
\end{figure}

In order to introduce relevant concepts and notations, we shall review
the formulation of the four-body problem in terms of a simpler
two-body problem \cite{Won04}.  We choose the four-body coordinate
system as shown in Fig.\ \ref{fig1} and label constituents $m_1^+$,
$m_2^-$, $m_3^{+}$, and $m_4^-$ as particles 1, 2, 3, and 4,
respectively with $m_1$$>$$m_2$.  The Hamiltonian for the
four-particle system is
\begin{eqnarray}
\label{eq1} H=\sum_{j=1}^4 \frac{ \bbox{p}_{j}^2} { 2 m_j} +
\sum_{j=1}^4 \sum_ {k>j}^4 v_{jk} +\sum_{j=1}^4 m_j,
\end{eqnarray}
in which particle $j$ has a momentum $\bbox{p}_j$ and a rest mass
$m_j$.  The pairwise interaction $v_{jk}(\bbox{r}_{jk})$ between
particle $j$ and particle $k$ depends on the relative coordinate
between them, $\bbox{r}_{jk}=\bbox{r}_j-\bbox{r}_k.$

We introduce the two-body total momentum $ {\bbox P}_{jk}=\bbox
{p}_j+\bbox{p}_k, $ and the two-body relative momentum $
\bbox{p}_{jk}=f_k \bbox{p}_j-f_j \bbox{p}_k, $ where 
\begin{eqnarray}
f_k&=&\frac{m_k}{m_{jk}},
\nonumber\\
m_{jk}&=&m_j+m_k. 
\end{eqnarray}
We choose to partition the four-body Hamiltonian in Eq.\ (\ref{eq1})
into an unperturbed $h_{12}+h_{34}$ for atoms $A(12)$ and ${\bar
  A}(34)$, a residual interaction $V_I$, and the kinetic energies of
$A$ and $\bar A$,
\begin{eqnarray}
H=h_{12} + h_{34}+V_I +\frac{ \bbox{P}_{12}^2}{2 m_{12}} + \frac{ \bbox{P}_{34}^2}{2
m_{34}} ,
\end{eqnarray}
where
\begin{eqnarray}
h_{jk}=\frac{ \bbox{p}_{jk}^2}{2 \mu_{jk}} +
v_{jk}(\bbox{r}_j-\bbox{r}_k) + m_{jk} {\rm
~~~~for~~}
\{jk\}=\{12\}{\rm~and~}\{34\},
\end{eqnarray}
and
\begin{eqnarray}
V_I=v_{14}(\bbox{r}_{14})+v_{13}(\bbox{r}_{13})
+   v_{23}(\bbox{r}_{23})+v_{24}(\bbox{r}_{24}).
\end{eqnarray} 
The eigenvalues of the unperturbed two-body Hamiltonians $h_{12}$ and
$h_{34}$ can be solved separately to obtain the bound state wave
functions and masses $M_{jk}(\nu)$ of atoms $A$ and ${\bar A }$,
\begin{eqnarray}
h_{jk}|(jk)_\nu\rangle =
[ \epsilon_{jk}(\nu) + m_{jk}] |(jk)_\nu\rangle
=M_{jk}(\nu)  |(jk)_\nu\rangle.
\end{eqnarray}
When we include $V_I$ as a perturbation, the eigenfunction
of ${H}$ becomes \cite{Lan58}
\begin{eqnarray}
\label{wave}
&\Psi&(\bbox{r},\bbox{r}_{12},\bbox{r}_{34})
=\psi(\bbox{r})\Biggl \{|A_0 {\bar A}_{0}\rangle
- {\sum_{\lambda,\lambda'}}' \frac{|A_{\lambda}
{\bar A}_{\lambda'}\rangle \langle A_{\lambda} {\bar A}_{\lambda'} | V_I | A_0
{\bar A}_{0} \rangle} {\epsilon_A({\lambda})+\epsilon_{\bar A}({\lambda'})-
\epsilon_A(0)-\epsilon_{\bar A}(0)} \Biggr \},~~~~~
\label{wf}
\end{eqnarray}
where $\bbox{r}= \bbox{R}_{12}-\bbox{R}_{34}$ is the interatomic
separation (see Fig.\ 1), $\bbox{R}_{jk} = f_j~\bbox{r}_{j}+
f_k~\bbox{r}_{k}$ is the center-of-mass coordinate of $m_j$ and $m_k$,
and $\sum_{\lambda \lambda'}'$ indicates that the sum is over a
complete set of atomic states $|A_{\lambda}{\bar
  A}_{\lambda'}\rangle$, including both bound and continuum states,
except $| A_{0} {\bar A}_{0}\rangle$.  The eigenvalue equation for the
four-body system with eigenenergy $\epsilon$ is
\begin{eqnarray}
\label{sch}
H\Psi (\bbox{r},\bbox{r}_{12},\bbox{r}_{34})
=[M_{12}(0)+M_{34}(0)+\epsilon]
 \Psi (\bbox{r},\bbox{r}_{12},\bbox{r}_{34}).
\end{eqnarray}
Working in the center-of-mass frame and taking the scalar product of
the above equation with $|A_{0} {\bar A}_{0}\rangle$, the eigenvalue
equation for the four-body system becomes the Schr\"odinger equation
for the motion of $A_{0}(12)$ relative to ${\bar A}_{0}(34)$,
\begin{eqnarray}
\label{sch1} \left \{\frac{\bbox{p}^2}{2 \mu}_{A{\bar A}} + V(\bbox{r})
\right \} \psi(\bbox{r}) =\epsilon \psi(\bbox{r}),
\end{eqnarray}
where $\bbox{p}$ is the relative momentum of the composite particles (atoms)
\begin{eqnarray}
\bbox{p}=
\frac {M_{34}(0)\bbox{P}_{12}-M_{12}(0)\bbox{P}_{34}}
      {M_{12}(0)+ M_{34}(0)},
\end{eqnarray}
 $\mu_{A\bar A}$ is the reduced mass of the two atoms
\begin{eqnarray}
\mu_{A{\bar A}}=\frac {M_{12}(0)M_{34}(0)}
          {M_{12}(0)+M_{34}(0)},
\end{eqnarray}
and  the interaction potential $V(\bbox{r})$
in Eq.\ (\ref{sch1}) is given by
\begin{eqnarray}
\label{VVV} 
V(\bbox{r})&=&\langle A_0 {\bar A}_{0} | V_I | A_0 {\bar A}_{0}
\rangle -{\sum}_{\lambda,\lambda'}'\frac{ |\langle A_{\lambda}
{\bar A}_{\lambda'} | V_I | A_0 {\bar A}_{0} \rangle|^2}
{\epsilon_A({\lambda})+\epsilon_{\bar A}({\lambda'})
-\epsilon_A({0})-\epsilon_{\bar A}({0})}.
\end{eqnarray}
We call the first leading-order term on the right hand side of the
above equation the direct potential, $V_{\rm dir}(\bbox{r})$,
\begin{eqnarray}
\label{vdir0}
V_{\rm dir}(\bbox{r})=\langle A_0 {\bar A}_{0} | V_I | A_0 {\bar A}_{0} \rangle,
\label{15}
\end{eqnarray}
which represents the Coulomb interaction between the constituents of
one atom and constituents of the other atom.  We call the second
next-to-leading order term in Eq.\ (\ref{VVV}) the polarization
potential, $V_{\rm pol}(\bbox{r})$,
\begin{eqnarray}
\label{vpoleq}
V_{\rm pol}(\bbox{r})=-{\sum}_{\lambda,\lambda'}'\frac{ |\langle A_{\lambda}
{\bar A}_{\lambda'} | V_I | A_0 {\bar A}_{0} \rangle|^2}
{\epsilon_A({\lambda})+\epsilon_{\bar A}({\lambda'})
-\epsilon_A({0})-\epsilon_{\bar A}({0})},
\label{16}
\end{eqnarray}
which is always negative.  It represents the effective interatomic
interaction arising from the virtual Coulomb excitation of the atoms
as they approach each other.

To study molecular states based on $A$ and $\bar A$ atoms as building
blocks, we quantize the Hamiltonian for the four-body system by
solving the Schr\" odinger equation (\ref{sch1}).  For such a purpose,
our first task is to evaluate the interaction potential $V(r)$ in
(\ref{VVV}) by calculating the direct and polarization potentials in
(\ref{15}) and (\ref{16}).

\begin{table}[h]
  \caption { The atomic length unit, Bohr radius, $a=\hbar/\alpha
    \mu$, and the atomic energy unit, $\alpha ^2\mu$, for different
    combinations of atomic constituents.}
\vspace*{0.0cm} 
\hspace*{1.5cm}
\begin{tabular}{|c|c|c|c|c|c|c|}
       \cline {3-7}
 \multicolumn{1}{c}{}   &\multicolumn{1}{c|}{}  &  $\mu^+$ & $\pi^+$  & K$^+$ &  p  & $\tau^+$    \\ \hline 
$e^-$& $a$ (fm)  & 53174 & 53111    & 52972   & 52946 & 52932\\
           &$\alpha^2 \mu $ (eV)  & 27.08 & 27.11 & 27.18 & 27.20 & 27.20 \\ \hline
$\mu^-$  &$a$ (fm)     &       & 449.7      & 310.7   & 284.7   & 271.1 \\ 
               & $ \alpha^2 \mu$ (keV) &   &  3.202  & 4.635      & 5.057   & 5.311\\ \hline 
$\pi^-$  &$a$ (fm)   &     &     & 248.5   & 222.6   & 209.0 \\ 
                 & $ \alpha^2 \mu$ (keV) &  &  &  5.794  & 6.47     & 6.891 \\ \hline 
$K^-$  &$a$ (fm) &     &        &   &  83.59   & 69.99 \\ 
              & $ \alpha^2 \mu$ (keV)  &  &   &   & 17.22     & 20.57\\ \hline 
$\bar p^-$  &$a$ (fm) &     &       &    &   & 44.04 \\ 
                        & $ \alpha^2 \mu$ (keV)  &  &   &    & & 32.69 \\ \hline 
\end{tabular}
\end{table}

For molecular states in the $A\bar A$ family, the atomic units of $A$
and $\bar A$ are the same.  To exhibit our results, it is convenient
to use the atomic unit of the $A(m_1^+ m_2^-)$ (or ${\bar A}$) atom as
our units of measurement:
\begin{enumerate}
\item All  lengths are measured in units of the Bohr radius 
of the $A(m_1^+ m_2^-)$ system, 
$ a_{12}={\hbar}{\alpha \mu_{12}}$,
where $\mu_{12}=m_1 m_2/(m_1+m_2)$.
\item All energy are measured in units of $\alpha ^2 \mu_{12}$ for the
  $A(m_1^+ m_2^-)$ system, which is two times the Rydberg energy, $2
  \epsilon_{{\rm Ryd}\{12\}}$.

\item
As a consequence, the reduced mass $\mu_{A\bar A} $ in the
Schr\"odinger equation (\ref{sch1}) for molecular motion in coordinate
${\bb r}$ needs to be measured in units of $\mu_{12}$,
\begin{eqnarray}
\label{18}
(\mu_{A{\bar A}} {\rm~ in~atomic~units}) = \frac{\mu_{A{\bar A}}}{\mu_{12}}
=\frac {M_{12}(0)M_{34}(0)}
          {\mu_{12}[M_{12}(0)+M_{34}(0)]}
\sim \frac{(m_1+m_2)^2}{2 m_1 m_2}.
\end{eqnarray}

\end{enumerate}
For brevity of notation, we shall omit the subscript $\{12\}$ in
$a_{12}$ and $ \alpha ^2 \mu_{12}$ except when it may be needed to
resolve ambiguities.  In Table I, we show the physical values of the
Bohr radius $a$ and the energy unit $\alpha^2\mu $.  They are
different for different $A$(and ${\bar A}$) atoms that build up the
$A\bar A$ molecule.  These quantities will be needed in Section X to
convert atomic units to physical units.

\section{  Method to Evaluate  $\langle A_{\lambda} {\bar A}_{\lambda'} | v_{jk}(r_{jk}) | A_0 {\bar A}_0 \rangle$}

To obtain the direct and polarization potentials, we need to evaluate
the matrix element $\langle A_{\lambda} {\bar A}_{\lambda'} |
V_{jk}(r_{jk}) | A_0 {\bar A}_0 \rangle$.  We shall examine first the
case when both $A_\lambda$ and ${\bar A}_{\lambda'}$ are bound states.
The case when one or both of $A_\lambda$ or ${\bar A}_{\lambda'}$ lies
in the continuum necessitates a different method and will be discussed
in Section VI.B.

When both $A_\lambda$ and ${\bar A}_{\lambda'}$ are bound, we can use
the Fourier transform method to evaluate the matrix elements
\cite{Sat83,Won04}.  Here, $A_\lambda (12)$ and ${\bar
  A}_{\lambda'}(34)$ can be represented by normalized hydrogen wave
functions $\phi_{\lambda}^{A}(\bbox{r}_{12})$ and
$\phi_{\lambda'}^{{\bar A}}(\bbox{r}_{34})$, respectively.  The
residual interaction $v_{jk}({\bb r}_{jk})$ is a function of ${\bb
  r}_{jk}$.  We need to express ${\bb r}_{jk}$ in terms of $\bbox{r}$,
${\bbox{r}}_{12}$, and ${\bbox{r}}_{34}$,
\begin{eqnarray}
{\bbox{r}}_{jk}={\bbox{r}}_{j}-{\bbox{r}}_{k} ={\bbox{r}}+F_A(jk)~{\bbox{r}}_{12}+F_{\bar A}(jk)~{\bbox{r}}_{34},
\label{eq6}
\end{eqnarray}
where the coefficients $F_{\{A,{\bar A}\}}(jk)$ have been given, with
a slight change of notations, in Ref.\ \cite{Won04},
\begin{eqnarray}
F_A(14)=~~f_2, &&~~ F_{\bar A}(14)=~~f_3,\nonumber\\ 
F_A(13)=~~f_2, &&~~ F_{\bar A}(13)=-f_4, \nonumber \\ 
F_A(23)=-f_1,  &&~~ F_{\bar A}(23)=-f_4,\nonumber\\
F_A(24)=-f_1,  &&~~ F_{\bar A}(24)=~~ f_3.\nonumber
\end{eqnarray}
 The matrix element $\langle A_\lambda
{\bar A}_{\lambda'}| v(\bbox{r}_{jk}) | A_0 {\bar A}_{0}\rangle$ can be written as
\begin{eqnarray}
\langle A_\lambda {\bar A}_{\lambda'}| v_{jk}(\bbox{r}_{jk}) |
A_0 {\bar A}_{0}\rangle
=\int d\bbox{r}_{12} d\bbox{r}_{34}
\rho_{\lambda 0}^A (\bbox{r}_{12}) \rho_{\lambda' 0}^{\bar A} (\bbox{r}_{34})
v_{jk}(\bbox{r}+F_A(jk)\bbox{r}_{12}+F_{\bar A}(jk)\bbox{r}_{34}),
\label{eq11}
\end{eqnarray}
where $\rho_{\lambda
0}^A(\bbox{r}_{12})$=$\phi_{\lambda}^*(\bbox{r}_{12})
\phi_0(\bbox{r}_{12})$ and  $\rho_{\lambda'
0}^{\bar A}(\bbox{r}_{34})$=$\phi_{\lambda'}^*(\bbox{r}_{34})
\phi_0(\bbox{r}_{34})$.
Introducing  the Fourier transform 
\begin{eqnarray}
{\tilde \rho}_{\lambda 0}^{A,{\bar A}}(\bbox{p}) =\int {d\bbox{y}} e^{i
\bbox{p}\cdot \bbox{y}} \rho_{\lambda 0}^{A,{\bar A}}(\bbox{y}),
\label{eq9}
\end{eqnarray}
and
\begin{eqnarray}
{\tilde v}_{jk}(\bbox{p}) =\int {d\bbox{r}_{jk}} e^{-i
\bbox{p}\cdot \bbox{r}_{jk}}
 v_{jk} (\bbox{r}_{jk}), \label{eq10}
\end{eqnarray}
we obtain
\begin{eqnarray}
\langle A_\lambda {\bar A}_{\lambda'}| v_{jk}(\bbox{r}_{jk}) | A_0
{\bar A}_{0}\rangle = \int \frac{d\bbox{p}}{(2\pi)^3} e^{i \bbox{p}\cdot
\bbox{r}} {\tilde \rho}_{\lambda 0}^A(F_A(jk)\bbox{p}) {\tilde
\rho}_{\lambda' 0}^{\bar A}(F_{\bar A}(jk)\bbox{p}) {\tilde v}_{jk}(\bbox{p}).
\label{eq22}
\end{eqnarray}
For our Coulomb potential
\begin{eqnarray}
{v}_{jk}({\bb r}_{jk}) = \frac{\alpha_{jk}}{|{\bb r}_{jk}|},
\label{eq12}
\end{eqnarray}
\begin{eqnarray}
\alpha_{jk}=\frac{e_j e_k}{\hbar  c},
\end{eqnarray}
where $e_j$ is the charge of $m_j$, 
 the Fourier transform of the Coulomb potential is
\begin{eqnarray}
{\tilde v}_{jk}(\bbox{p}) =\frac{4\pi \alpha_{jk}}{{\bb p}^2}.
\end{eqnarray}
The Fourier transform ${\tilde \rho}_{\lambda 0}^A(F_A(jk)\bbox{p})$ depends on the sign of $F_A(jk)$ and the $l$ quantum number of 
$|A_\lambda\rangle = |nlm\rangle $.    
 It is easy to show that 
\begin{eqnarray}
 {\tilde \rho}_{\lambda 0}^A(F_A(jk)\bbox{p})
= [{\rm sign}(F_A(jk))]^l~{\tilde \rho}_{\lambda 0}^A(f_A(jk)\bbox{p}),
\label{26}
\end{eqnarray}
where ${\rm sign}(F_A(jk))$ is the sign of
$F_{A}(jk)$, and $f_{A}(jk)$ is the magnitude of $F_{A}(jk)$,
\begin{eqnarray}
f_A(jk)=|F_A(jk)|.
\end{eqnarray}
Substituting Eq.\ (\ref{26}) in Eq.\ (\ref{eq22}), we obtain
\begin{eqnarray}
\label{30}
\langle A_\lambda {\bar A}_{\lambda'}| v_{jk}(\bbox{r}_{jk}) | A_0
{\bar A}_{0}\rangle =s \int \frac{d\bbox{p}}{(2\pi)^3} e^{i \bbox{p}\cdot
\bbox{r}} {\tilde \rho}_{\lambda 0}^A(f_A(jk)\bbox{p}) {\tilde
\rho}_{\lambda' 0}^{\bar A}(f_{\bar A}(jk)\bbox{p}) {\tilde v}_{jk}(\bbox{p}),
\end{eqnarray}
where $s$ is the sign factor
\begin{eqnarray}
s=[{\rm sign}F_A(jk)]^l~ [{\rm sign}F_{\bar A}(jk)]^l.
\label{eqs}
\end{eqnarray}

\section{The Direct Potential $V_{\rm dir}({\bb r})$}

\begin{figure} [h]
\hspace*{0.0cm}
\includegraphics[scale=0.45]{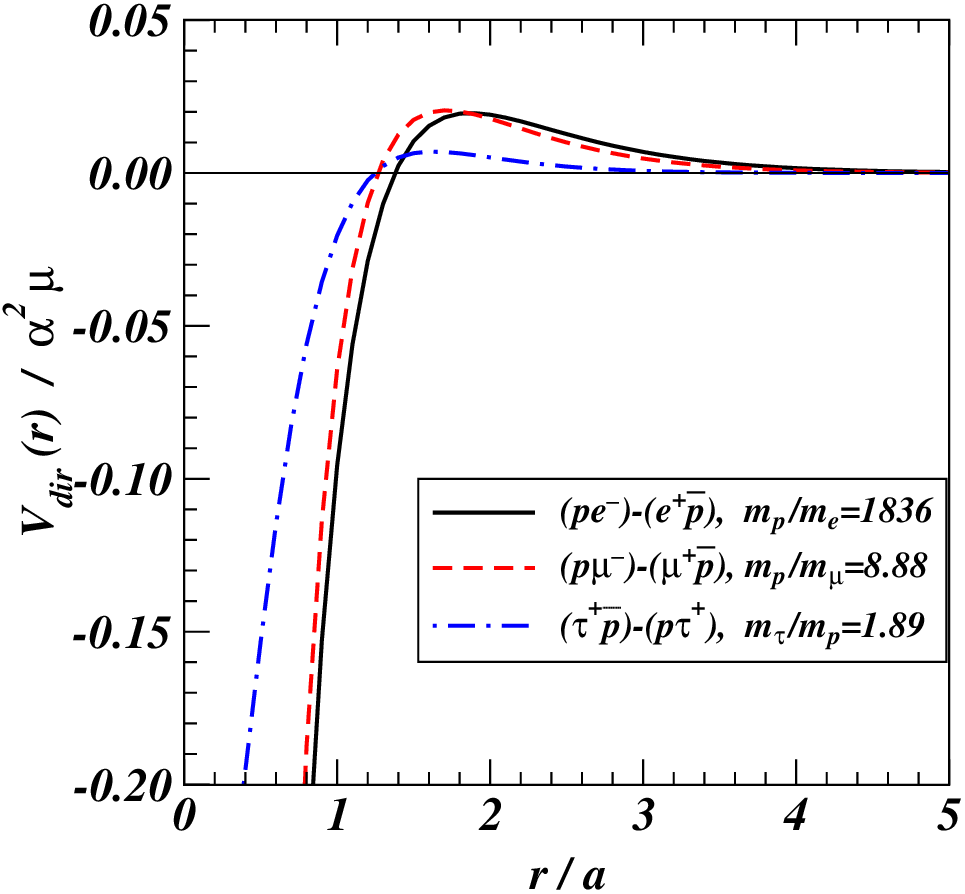}
\caption{ The direct potential $V_{\rm dir}(r)$ in atomic units, for
  different systems with different $m_1/m_2$. }
\end{figure}

The direct potential is equal to the matrix element $\langle A_0 {\bar
  A }_0|V_I|A_0 {\bar A}_0\rangle$ .  We can apply the results of
Eq.\ (\ref{30}) to evaluate this matrix element.  We note that
\begin{eqnarray}
{\tilde \rho}_{00}^A ({\bf p})
=   \frac {16 } {((pa)^2+4)^2} .
\label{eq92}
\end{eqnarray}
Substituting this into Eq.\ (\ref{30}), we have
\begin{eqnarray}
&&\langle A_{0} {\bar A}_{0}| v_{jk}(\bbox{r}_{jk}) | A_{0} {\bar A}_{0}\rangle
= \int \frac{d\bbox{p}}{(2\pi)^3} e^{i \bbox{p}\cdot
\bbox{r}}
\frac {16 } {((pf_A a)^2+4)^2}   \frac {16 } {((pf_{\bar A} a_{34})^2+4)^2}  \frac{4\pi\alpha_{jk}}{p^2},~~~~~~
\label{32}
 \end{eqnarray}
which leads to the direct potential 
\begin{eqnarray}
V_{\rm dir}(r)=&&
 \frac{\alpha e^{-2 r/f_1 a}}{r} 
\biggl  \{
\left ( -\frac{2f_1^4(f_1^2-3 f_2^2)}{(f_1^2 -f_2^2 )^3} +1 \right )
+\left (-\frac{2f_1^4}{(f_1^2 -f_2^2 )^2} + 1\right ) \left (\frac{ r}{f_1 a}\right )
\nonumber\\
&&
+\frac{1}{4}\frac{r}{f_1 a}
\left ( 1+2\frac{r}{f_1 a} \right )
+\frac{1}{24}\frac{r}{f_1 a}
\left [4\left ( \frac{r}{f_1 a}\right )^2  
+ 6\frac{r}{f_1 a}+3\right ] \biggr \}
\nonumber\\
&&
 + \frac{\alpha e^{-2 r/f_2 a}}{r} 
\biggl  \{\left ( -
 \frac{2f_2^4(f_2^2-3 f_1^2) }{(f_2^2 -f_1^2 )^3}   +1\right )
+\left ( - \frac{2f_2^4}{(f_2^2 -f_1^2 )^2} +1 \right ) \left (\frac{ r}{f_2 a}\right ) 
\nonumber\\
&&
+\frac{1}{4}\frac{r}{f_2 a}
\left ( 1+2\frac{r}{f_2 a} \right )
+\frac{1}{24}\frac{r}{f_2 a}
\left [4\left( \frac{r}{f_2 a}\right)^2  
+ 6\frac{r}{f_2 a}+3\right ] \biggr \}.~~~~~~~~
\label{dir}
\end{eqnarray}
This direct potential is a sum of two Yukawa potentials of screening
lengths $f_1 a/2$ and $f_2 a/2$, multiplied by third-order polynomials
in $r$.  It can be shown numerically or analytically that for
$m_1=m_2$, the direct potential $V_{\rm dir}$ is zero.

In the region close to $r\to 0$, the direct potential becomes
\begin{eqnarray}
\lim_{r\to 0}
V_{\rm dir}(r)=&&
 \frac{\alpha e^{-2 r/f_1 a}}{r} 
\left ( -\frac{2f_1^4(f_1^2-3
   f_2^2)}{(f_1^2 -f_2^2 )^3} +1 \right ) 
+
\frac{\alpha e^{-2 r/f_1 a}}{f_1 a}
\left (-\frac{2f_1^4}{(f_1^2 -f_2^2 )^2} +
 \frac{5}{4} \right )
\nonumber\\
+&& \frac{\alpha e^{-2 r/f_2 a}}{r} 
\left ( - \frac{2f_2^4(f_2^2-3 f_1^2)}{(f_2^2 -f_1^2 )^3}
 +1 \right )
 +\frac {\alpha e^{-2 r/f_2 a}}{f_2 a}
\left ( -
 \frac{2f_2^4}{(f_2^2 -f_1^2 )^2} +\frac{5}{4} \right ).~~~~~~~
\end{eqnarray}
If $m_1 \gg m_2$ in this  region close to $r\to 0$, then  
the direct potential   becomes
\begin{eqnarray}
\lim_{r\to 0, ~m_1 \gg m_2}
V_{\rm dir}(r) \sim 
- \frac{\alpha e^{-2 r/f_1 a}}{r} + \frac{\alpha e^{-2 r/f_2 a}}{r}
-
\frac{3\alpha }{4f_1 a}
 +\frac {5\alpha }{4f_2 a} .
\label{35}
\end{eqnarray}
For this case of $m_1 \gg m_2$, we have $f_1 a/2$$\sim$$a_{m_1 m_2}/2$
and $f_2 a/2$$\sim$$ a_{m_1 {\bar m}_1}/4$, where $a_{m_i m_j}$=$
(m_i$+$m_j)/\alpha$$m_i m_j$ is the Bohr radius of the $(m_i m_j)$
system.  The first term is a screened Coulomb interaction with the
range of $a_{m_1 m_2}/2$ and the second term is a repulsive screened
Coulomb potential with a range of $ a_{m_1 {\bar m}_1}/4$.  The last
three terms reduce the binding energies of molecular bound states in
the case of $m_1 \gg m_2$.

Equation (\ref{dir}) is a general result applicable to any mass ratio
of $m_1/ m_2$, and is a generalization of the results of \cite{Mor73}
that represents only an approximation for $m_1/m_2 \gg 1$.

We show in Fig.\ 2 the direct potential in atomic energy units,
$\alpha^2 \mu$, for $(pe^-)$-$(e^+{\bar p})$, $(p\mu^-)$-$(\mu^+{\bar
  p})$, and $(\tau^+{\bar p})$-$(p\tau^-)$, as a function of the
interatomic separation $r$ in atomic units, $a$.  For systems with a
large ratio of $m_1/m_2$, the interaction is slightly repulsive at
large separations, owing to the repulsion of like charges.  The
repulsive interaction of like charges is strongest when the atoms are
nearly ``touching" each other at $r\sim 2 a$, leading to development
of a barrier there.  At $r < a$, the interaction between the heavy
unlike charges dominates, and the direct potential changes to become
strongly attractive.  We observe in Fig.\ 2 that as the mass ratio
$m_1/m_2$ approaches unity, there is a cancellation of both the
attractive and repulsive components of the direct potential.  The
repulsive barrier is lowered and the direct potential becomes less
attractive at short distances.  In fact, as we noted earlier, $V_{\rm
  dir}$ vanishes if $m_1=m_2$.

 \section{The Polarization Potential $V_{\rm pol}({\bb r})$}

The polarization potential $V_{ \rm pol}(r)$ is the effective
interaction between $A$ and $\bar A$ arising from virtual atomic
excitations.  It can be obtained as a double summation over
$A_\lambda$ and ${\bar A}_{\lambda'}$ involving the excitation matrix
element $\langle A_\lambda {\bar A}_{\lambda'}| V_{jk}(\bbox{r}_{jk})
| A_0 {\bar A}_{0}\rangle$.  The summation over $A_\lambda$ and ${\bar
  A}_{\lambda'}$ states includes the complete set of bound and
continuum states, but excludes the ground states.  We shall make the
assumption that the virtual excitation is predominantly electric
dipole in nature and shall truncate the set of excited $A_\lambda$ and
$A_{\lambda'}$ states to include only $l=1$ states.  Because the
ground states $A_0$ and ${\bar A}_0$ have no orbital angular momentum,
the azimuthal quantum numbers $m$ of $A_\lambda$ and ${\bar
  A}_{\lambda '}$ must be equal and opposite.  The polarization
excitation therefore contains contributions where $\lambda$-$\lambda'$
are bound-bound (bb), bound-continuum (bc), and continuum-continuum
(cc), with $l$=$1$.  We shall discuss separately how these excitation
matrix elements can be evaluated.

\subsection{ Bound-bound excitation matrix elements} 
When both $ A_\lambda$ and ${\bar A}_{\lambda'} $ are bound states,
the ``bound-bound" excitation matrix element $\langle A_\lambda {\bar
  A}_{\lambda'} |V_I|A_0 B_0\rangle$ can be evaluated using the method
of Fourier transform.  The results were presented previously in
\cite{Lee08}.  We shall rewrite the same result in a slightly
simplified form.  As shown in Appendix A, the relevant matrix element
for $|A_\lambda\rangle =|n l m\rangle$ and $|{\bar
  A}_{\lambda'}\rangle =|n'l~ (-m)\rangle$ with $l=1$ is given by
\begin{eqnarray}
\label{mat}
\langle A_\lambda B_{\lambda'} |v_{jk}({\bb r}_{jk})|A_0 B_0\rangle
= s
\int \frac{p^2 dp}{(2\pi)^3} & 
{\tilde R}_{n 1 m,100}^A[f_A(jk)p]  ~
{\tilde R}_{n'1\, (-m),100}^{\bar A}[f_{\bar A}(jk)p] 
\nonumber\\
&\times
{\tilde v}_{jk}(p)  J(p,r),
\end{eqnarray}
where $s$ is the sign factor as given by (\ref{eqs}), $J(p,r)$ is
given in terms of the spherical Bessel functions,
\begin{eqnarray}
J(p,r)=
\begin{cases}
 j_0(pr) -2j_2(pr)& ~~~{\rm for~} m=0, \cr
-[j_0(pr)+ j_2(pr)] & ~~~{\rm for~} m=1, \cr
 \end{cases}
\end{eqnarray}
${\tilde R}_{n1m,100} (p)$ is given in terms of the Genegbauer
polynomial $C_\mu^\nu$,
\begin{eqnarray}
&&{\tilde R}_{n1m,100} (p)
=\frac{\sqrt{4\pi}N_{10}N_{n1}}{n+1}  \left ( \frac{na}{2}\right )^3 
 \sum_{\kappa=0}^{n-2} \left ( \begin{matrix}
n-2+3\cr
  \kappa 
\end{matrix} \right )\beta^{n-2-\kappa} (1-\beta)^\kappa
 \nonumber\\
&&\times 
\frac{{npa (n+1)^{2}~ 2^6(n-\kappa) }}{((npa)^2+(n+1)^2)^3} 
C_{n-2-\kappa}^{2} \left (\frac{(npa)^2-(n+1)^2}{(npa)^2+(n+1)^2}\right ),
\label{rr}
\end{eqnarray}
the variable $\beta$ is
\begin{eqnarray}
\beta=\frac{1}{n+1},
\end{eqnarray}•
and the normalization constant $N_{nl}$ is 
\begin{eqnarray}
N_{nl}=\sqrt{ \frac{ 4 (n-l-1)!}{a^3n^4[(n+l)!]}} .
\end{eqnarray}
Thus, the six dimensional integral over ${\bb r}_{12}$ and ${\bb
  r}_{34}$ in the matrix element is reduced into a one-dimensional
integral that can be readily carried out numerically.

\subsection{Bound-continuum and continuum-continuum excitation matrix elements}

For a given interatomic separation $r$, the excitation matrix element
 $\langle A_\lambda {\bar  A}_{\lambda'} |V_I|A_0 B_0\rangle$ involving one or two
continuum states can be evaluated by direct numerical integration in
the six-dimensional space of ${\bb r}_{12}$ and ${\bb r}_{34}$.  As
constrained by the ground state wave functions of $A_0$ and ${\bar
  A}_0$, the integrand in such an integration has weights concentrated
around the region of ${\bb r}_{12}\sim 0$ and ${\bb r}_{34}\sim 0$,
and the continuum wave function does not need to extend to very large
distances.

Following Bethe and Salpeter \cite{Bet57}, we use the radial wave
function for a continuum state of $A$ (or ${\bar A}$) with momentum
$|{\bb k}|=k$ as given by
\begin{eqnarray}
R_{kl}(r)=\frac {1}{kr} \sqrt{\frac {2}{\pi}} F_l(\eta,kr), 
\end{eqnarray}
where $F_l(\eta,kr)$ is the regular Coulomb wave function
\cite{Abr70,Bar81}.  The coefficient of the wave function has been
chosen according to the normalization
\begin{eqnarray}
\sum_{l m} \int k^2 dk\left  |\frac {1}{kr} \sqrt{\frac {2}{\pi}} F_l(\eta,kr) Y_{l m}(\theta, \phi) \right \rangle
\left  \langle \frac {1}{kr} \sqrt{\frac{ 2}{\pi}} F_l(\eta,kr) Y_{l m}(\theta, \phi)\right | =1.
\end{eqnarray}
For the excitation to a continuum state in ${ A}$ and a bound state in
$\bar A$ this closure relation allows us to write the bound-continuum
contribution to the polarization potential to be
\begin{eqnarray}
V_{\rm pol}^{bc} (\bbox{r})=-\sum_{lm} \int k^2 dk{\sum}_{\lambda'}'
\frac{ |\langle 
R_{k l}^{A}({ r}_{12}) Y_{lm} (\theta_{12}\phi_{12}){\bar A}_{\lambda'}| V_I | A_0 {\bar A}_{0} \rangle|^2}
{\epsilon_A({k})+\epsilon_{\bar A}(\lambda')
-\epsilon_A({0})-\epsilon_{\bar A}({0})}.
\label{44}
\end{eqnarray}
There is a similar contribution for the excitation into a bound state
in $A$ and a continuum state in ${\bar A}$.

The continuum-continuum contribution to the polarization $V_{\rm
  pol}^{cc}({\bb r})$ is given similarly by
 \begin{eqnarray}
&&V_{\rm pol}^{cc} (\bbox{r})=-\sum_{lm}\int k^2 dk \sum_{l'} \int k'^2 dk'
\nonumber\\
&&\times
\frac{ |\langle R_{k l}^{A}({ r}_{12}) Y_{lm} (\theta_{12}\phi_{12})
R_{k'l'}^{\bar A}({ r}_{34}) Y_{l'-m} (\theta_{34}\phi_{34})| V_I | A_0 {\bar A}_{0} \rangle|^2}
{\epsilon_A({k})+\epsilon_{\bar A}(k')
-\epsilon_A({0})-\epsilon_{\bar A}({0})}.
\label{45}
\end{eqnarray}
To evaluate the excitation matrix element of $V_I$ in Eqs.\ (\ref{44})
and (\ref{45}), we discretized the continuum momentum $k$ (and $k'$)
into momentum bins. For each of the bins, the wave functions in terms
of the the spatial coordinates ${\bb r}_{12}$ and ${\bb r}_{34}$ are
all known.  We shall again limit our consideration to dipole
excitations with $l=l'=1$ only.  In the numerical calculations, the
residual interaction $V_I$ is a sum of four Coulomb interactions which
depend on the magnitude of the radius vector ${\bb r}_{jk}$ where
$\{jk\}=\{14\}$, $\{13\}$, $\{23\}$, and $\{24\}$.  The square of the
magnitude $|{\bb r}_{jk}|^2$ can be evaluated as
\begin{eqnarray}
&|{\bb r}_{jk}|^2&
=
 r^2 + r_{12}^2 + r_{34}^2 + 2 F_A(jk) r r_{12} \cos \Omega({\bb r}, {\bb r}_{12}) 
\nonumber\\
& &+ 2 F_B(jk) r r_{34} \cos \Omega({\bb r}, {\bb r}_{34}) 
 +2 F_A(jk)  F_B(jk) r_{12} r_{34} \cos \Omega({\bb r}_{12}, {\bb r}_{34}).~~~~~~
\label{eq15}
\end{eqnarray}
It is convenient to choose the coordinate systems of ${\bb r}$, ${\bb
  r}_{12}$, and ${\bb r}_{34}$ such that their $z$-axes lie in the
same direction, and their corresponding $x$- and $y$-axes are parallel
to each other.  With this choice of the axes, we have
\begin{eqnarray}
\cos \Omega({\bb r}, {\bb r}_{12}) &=&\cos \theta_{12},\\
\cos \Omega({\bb r}, {\bb r}_{34}) &=&\cos \theta_{34},\\
\cos \Omega({\bb r}_{12}, {\bb r}_{34}) 
&=&\cos \theta_{12}\cos \theta_{34}
+\sin \theta_{12} \sin \theta_{34} \cos (\phi_{12}-\phi_{34})  .
\end{eqnarray}
These relations allow us to determine the integrand and carry out the
6-dimensional integration in ${\bb r}_{12}$ and ${\bb r}_{34}$, for the
evaluation of the excitation matrix element and the polarization
potential.

\section{The interaction potential in $(pe^-)$-$(e^+{\bar p})$}

\begin{figure} [h]
\hspace*{0.0cm} \includegraphics[scale=0.45]{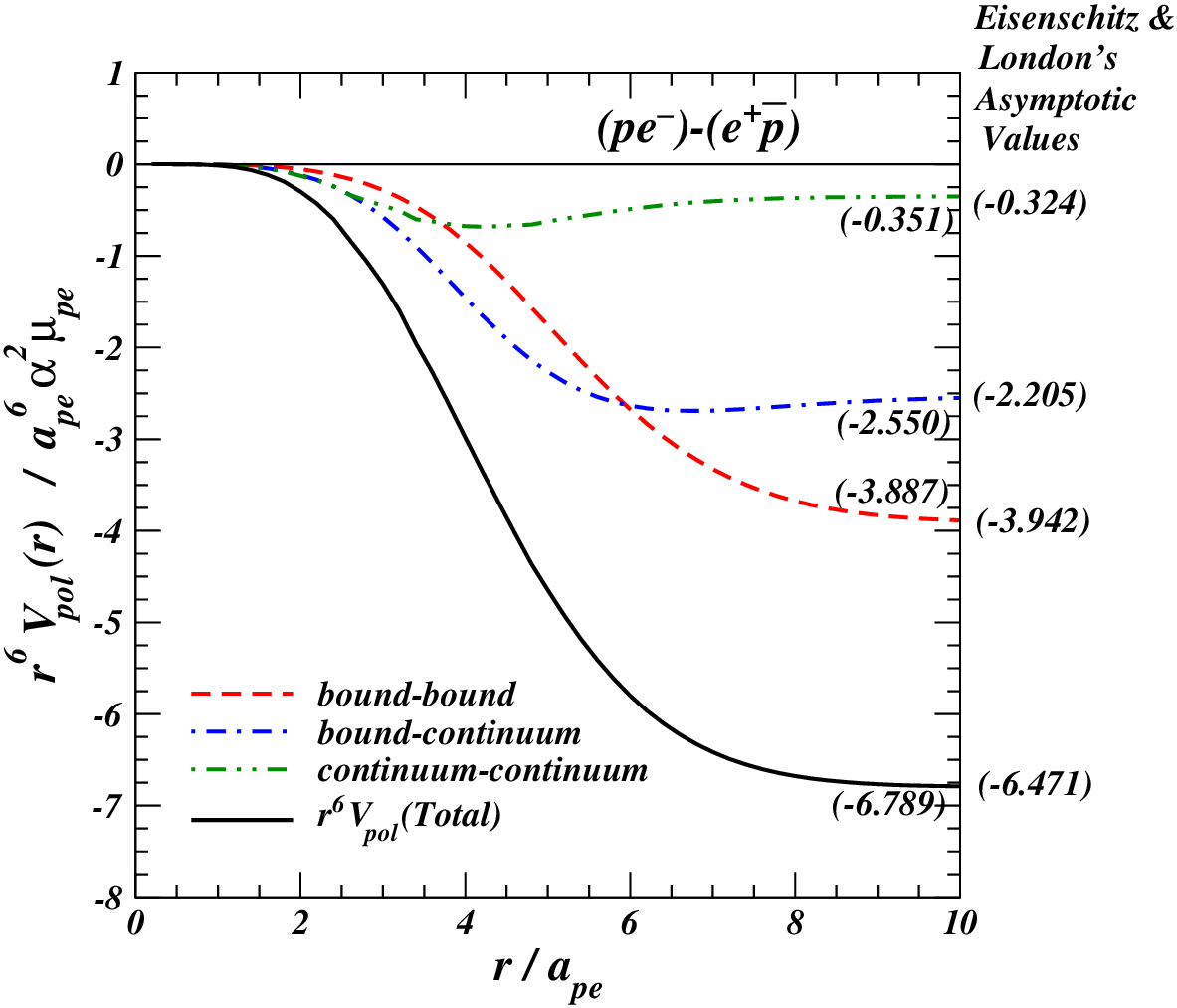}
\caption{ The quantity $r^6 V_{\rm pol}(r)$ for different polarization
  potential components plotted as a function of $r$ in atomic units,
  for the $(pe^-)$-$(e^+{\bar p})$ system.  The numbers on the right
  are the $C_6$ coefficients from Eisenshitz and London for H${}_2$
  \cite{Eis28}.  }
\end{figure}

Using the methods discussed in the last section, excitation matrix
elements can be evaluated and the polarization potential $V_{\rm pol}$
from different contributions can be obtained.  We show in Fig.\ 3 the
quantity $r^6 V_{\rm pol}(r)$ for the $(pe^-)$-$(e^+{\bar p})$ system
as a function of $r$, where $r^6 V_{\rm pol}(r)$ from the bound-bound,
bound-continuum, and continuum-continuum contributions are shown as
the dashed, dash-dot and dash-dot-dot curves respectively.  In these
calculations, we include bound states up to $n=20$ in bound-bound
calculations and $n=12$ in bound-continuum calculations.  For
calculations with continuum states, we include states up to $k=6$
atomic units.

We note in Fig.\ 3 that the curves of $r^6V_{\rm pol}$ flatten out at
large values of $r$.  This indicates that the attractive polarization
potentials behave asymptotically as $-C_6/r^6$, the well-known van der
Waals interaction at large distances between atoms.  The asymptotic
values of the $C_6$ from different contributions have been given along
with the theoretical curves.  They have also been obtained previously
by Eisenschitz and London for H${}_2$ \cite{Eis28} as listed on the
right side of the figure.  There is reasonable agreement between the
$C_6$ values obtained in the present calculation and those of
\cite{Eis28}.  There is a small difference between the $C_6$ numbers
involving continuum states with those of \cite{Eis28}.  These small
differences may arise from the fact that the curves involving
continuum states have not yet become completely flattened and thus
they may have not yet reached their asymptotic values.

\begin{figure} [h]
\hspace*{0.0cm}
\includegraphics[scale=0.45]{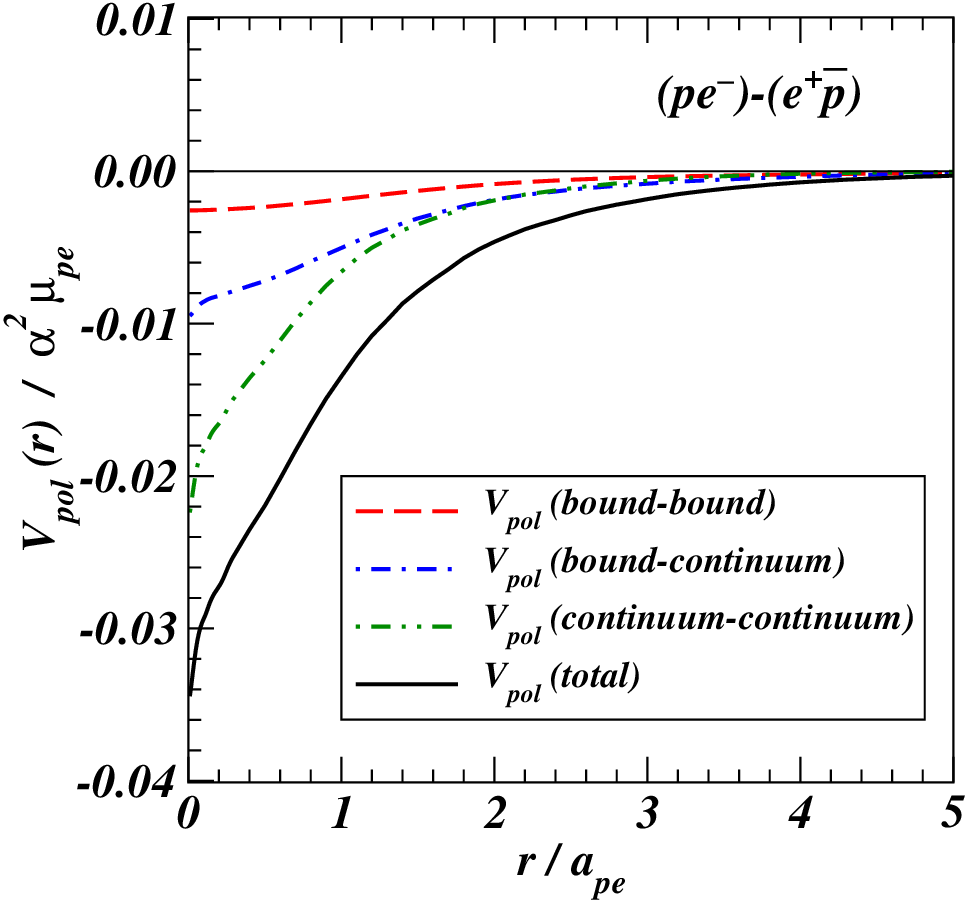}
\caption{ Components of the  polarization potential  and their total sum for
the  $(p e^-)$-$(e^+{\bar p})$ system. } 
\end{figure}

From these results, we note that at large separations, the
bound-continuum contribution is much larger than the
continuum-continuum contribution and is slightly smaller than the
bound-bound contribution.
 
\begin{figure} [h]
\hspace*{0.0cm}
\includegraphics[scale=0.45]{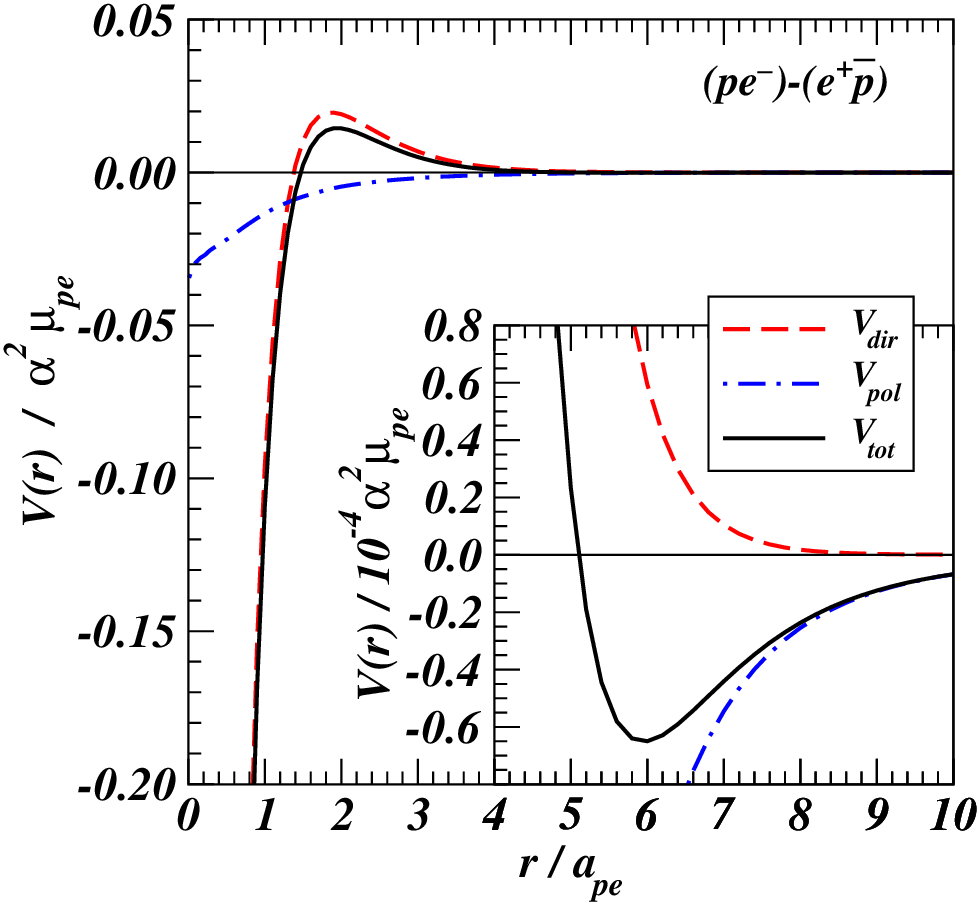}
\caption{ The direct, polarization and total interaction potentials
  for the $(p e^-)$-$(e^+{\bar p})$ system. }
\end{figure}

The situation is different at small separations.  We plot $V_{\rm
  pol}({ r})$ as a function of $r$ in Fig.\ 4 for $(pe^-)$-$(e^+\bar
p)$ and $r<5 a_{pe}$.  We find that the bound-bound contributions are
smaller than the bound-continuum contributions, which in turn are
smaller than the continuum-continuum contributions.  The total
polarization potential remains well-behaved at small $r$.  Its
magnitude is much smaller than the magnitude of the direct potential
dominated by the attractive screened potential between $p$ and $\bar
p$.

Having obtained both the direct and the total polarization potential,
we can add them together to obtain the interaction potential
$V(r)=V_{\rm dir}(r)+V_{\rm pol}(r)$.  We show in Fig.\ 5 the
interaction potential $V(r)$ and its components $V_{\rm dir}(r)$ and
$V_{\rm pol}(r)$ for the $(pe^-)$-$(e^+\bar p)$ system.  We note that
for this case of large ratio of constituent masses $m_1/m_2$, the
polarization potential is small compared to the direct potential at
short distances and the total interaction is attractive at $r< a$.
The repulsive barrier at $r\sim 2 a$ that comes from the direct
potential remains.  The repulsive interaction decreases at larger
separations.  There is a pocket structure at $r\sim 6 a_{pe}$ that is
very shallow and arises from the interplay between the repulsion of
like charges at intermediate distances and the attractive polarization
potential \cite{Lee08}.

\section{The interaction potential in different systems}

It is of interest to see how the interaction potential and its various
components vary as a function of the constituent masses.  The various
potential components and the total polarization potential for
$(\mu^+e^-)$-$(e^+ \mu^-)$ in atomic units are very similar to those
of $(pe^-)$-$(e^+\bar p)$ and will not be presented.  The situation
changes slightly for $(p \mu^-)$-$(\mu^+\bar p)$.  We show the
polarization potential components in Fig.\ 6 and the total interaction
potential in Fig.\ 7 for $(p \mu^-)$-$(\mu^+\bar p)$.  One notes from
Fig.\  6 that for this case with $m_p/m_\mu = 8.88$, the bound-bound
contributions to the polarization potential dominate over the
bound-continuum or the continuum-continuum contributions.  The results
in Fig.\ 7 indicate that the total polarization potential is however
small compared to the direct potential at $r<a$.  The other features
of the interaction potential is similar to those of the $(p
e^-)$-$(e^+{\bar p})$ system.

\begin{figure} [h]
\hspace*{0.0cm}
\includegraphics[scale=0.45]{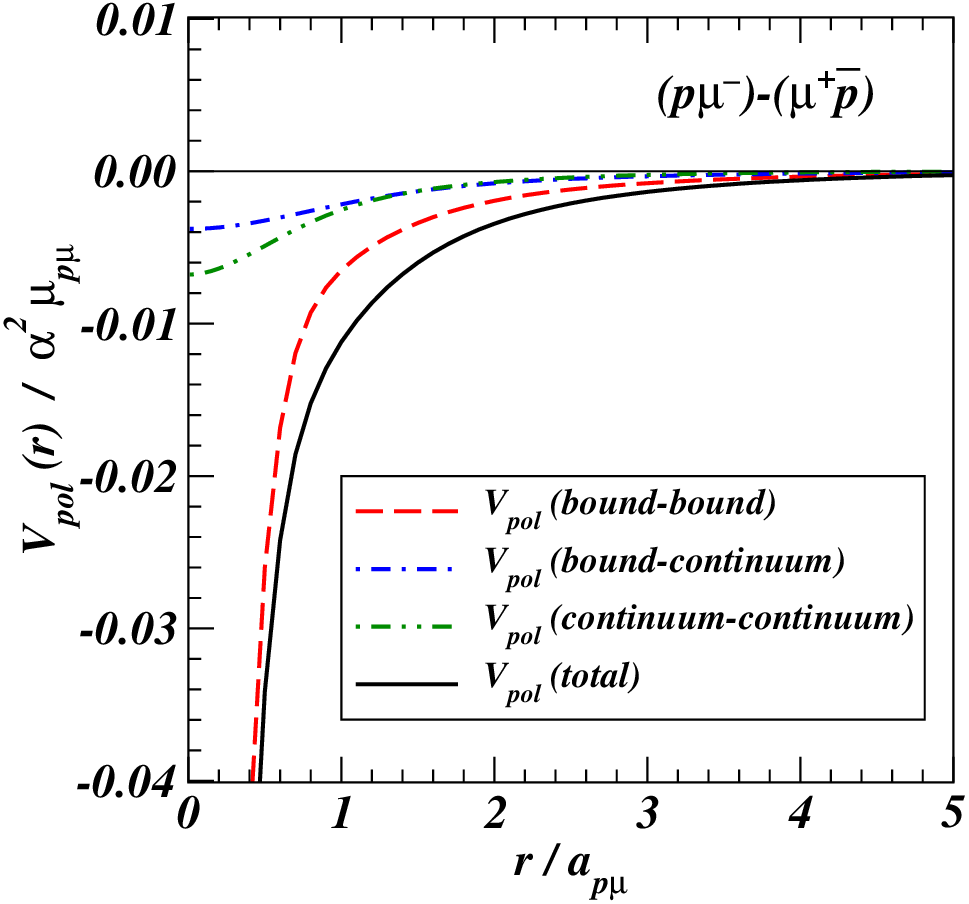}
\caption{ 
Components of the  polarization potential and their total sum 
for the  $(p\mu^-)$-$(\mu^+{\bar p})$ system. 
}
\end{figure}

\begin{figure} [h]
\hspace*{0.0cm}
\includegraphics[scale=0.45]{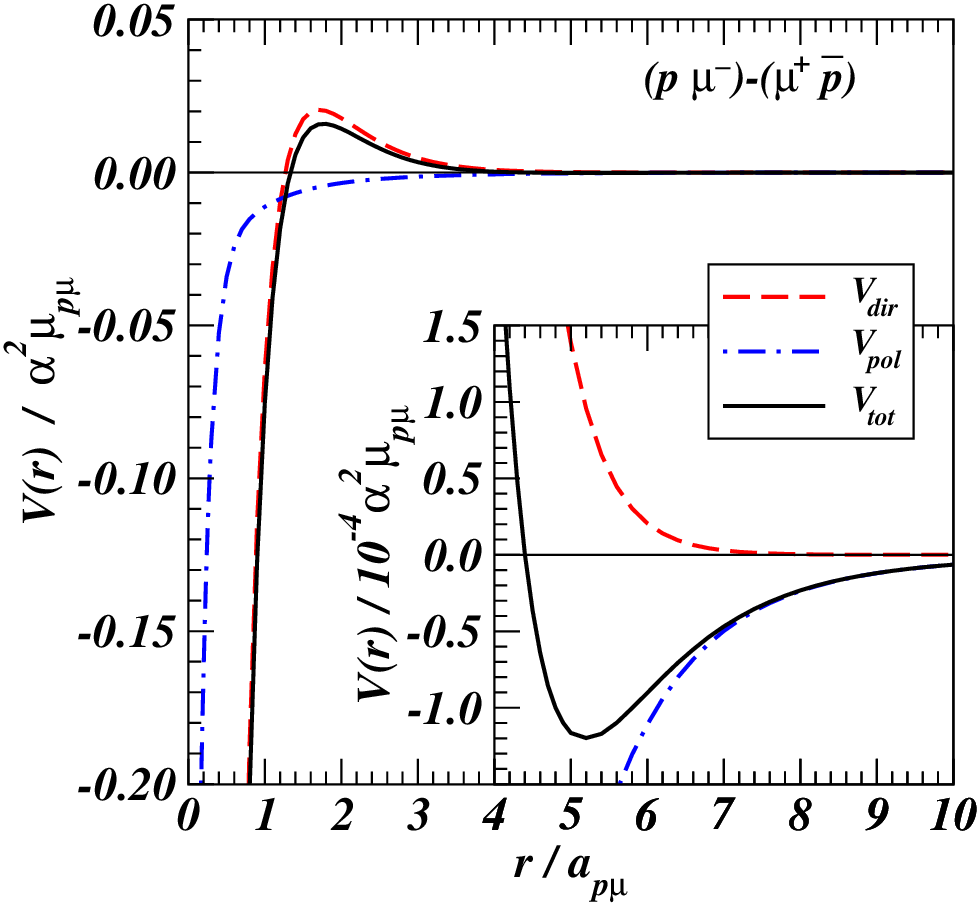}
\caption{ The direct potential, polarization potential, and the total
  interaction potential for the $(p\mu^- )$-$(\mu^+{\bar p})$ system. }
\end{figure}

When the ratio of constituent masses $m_1/m_2$ approaches unity, the
qualitative features of the different components change significantly.
In Fig.\ 8, we show various components of the interaction potential for
the $(\tau^+ {\bar p})$-$(p\tau^-)$ system, for which
$m_\tau/m_p=1.89$.  As one observes, the polarization potential is
dominated by the bound-bound component while the bound-continuum and
continuum-continuum contributions are small.  The direct potential is
much reduced compared to the case with large constituent mass ratios
$m_1/m_2$ and is now of the same order of magnitude as the
polarization potential.

\begin{figure} [h]
\hspace*{0.0cm}
\includegraphics[scale=0.45]{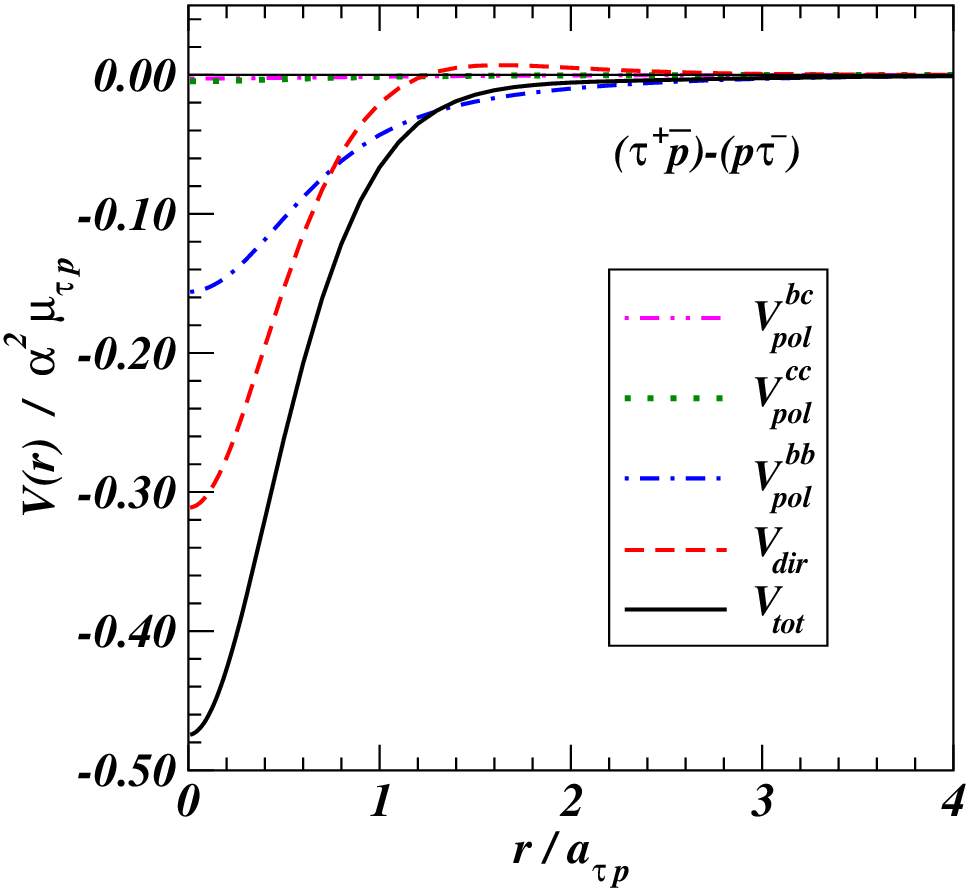}
\caption{ The direct potential, polarization potential, and the total
  potential for the $(\tau^+{\bar p})$-$(p\tau^-)$ system. }
\end{figure}

\section{Quantization of the four-body system}
\begin{figure} [h]
\hspace*{0.0cm}
\includegraphics[scale=0.45]{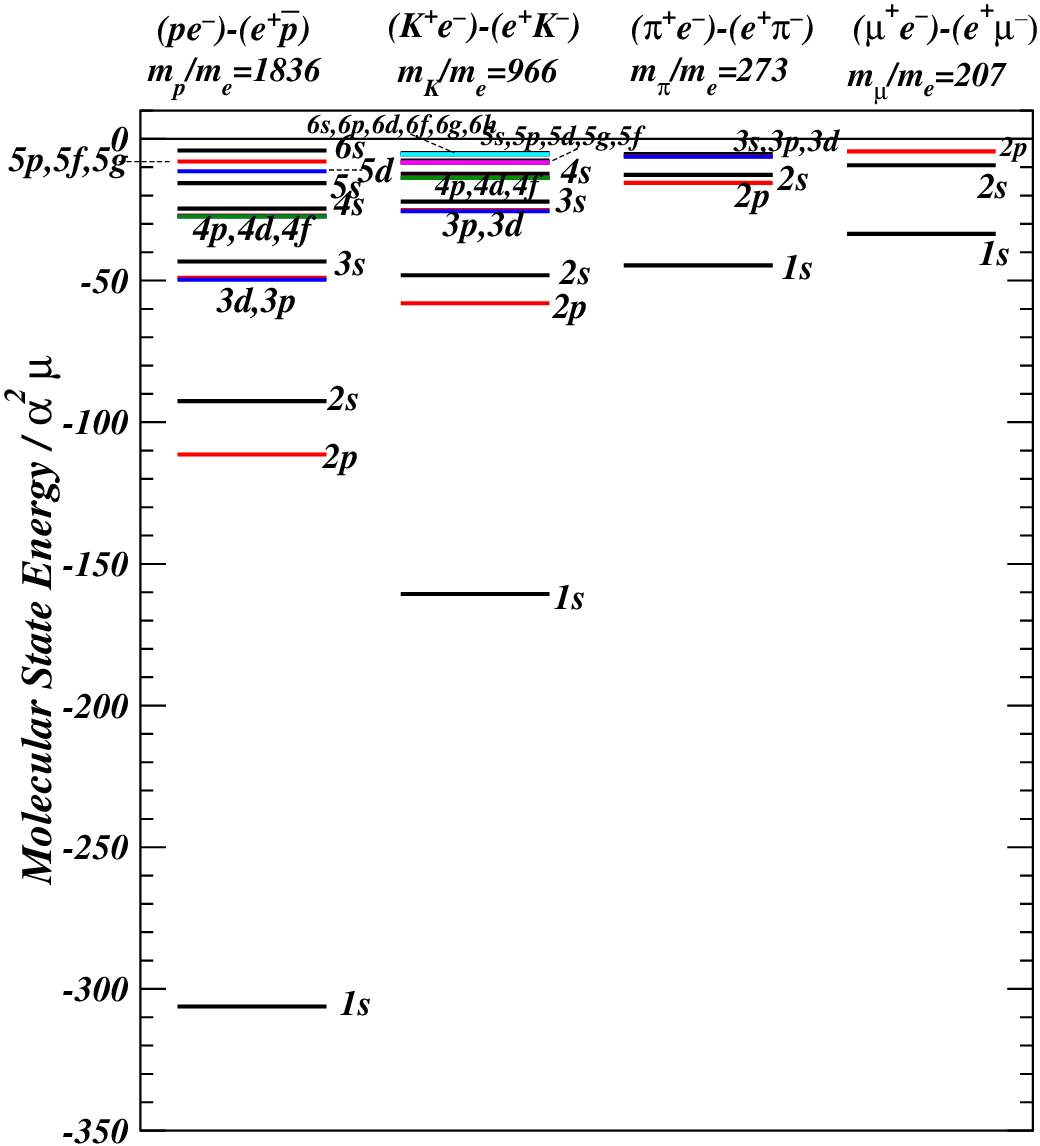}
\caption{ The eigenenergies of the molecular states states in $(pe^-)$-$(e^+{\bar p})$,
$(K^+ e^-)$-$(e^+K^-)$, $(\pi^+ e^-)$-$(e^+\pi^-)$, and $(\mu^+e^-)$-$(e^+\mu^-)$,
in atomic units.
}
\end{figure}

To study molecular states based on $A$ and $\bar A$ atoms as building
blocks, we quantize the Hamiltonian for the four-body system by
solving the Schr\" odinger equation (\ref{sch1}).  The states of the
system depend not only on the interaction potential $V(r)$ but also on
the reduced mass.

Because we use the atomic units of the $A(m_1 m_2)$ atom to measure
our quantities, it is necessary to measure the reduced mass
$\mu_{A{\bar A}}$ for molecular motion in units of $\mu_{12}$, as
given by Eq.\ (\ref{18}),
\begin{eqnarray}
(\mu_{A{\bar A}} {\rm~ in~atomic~units}) 
=\frac{(m_1+m_2)^2}{2 m_1 m_2}
=\frac{(1+m_1/m_2)^2}{2 m_1 /m_2}.
\end{eqnarray}
To provide a definite description of the constituents, we order the
masses of the constituents such that $m_1$$>$$m_2$ and characterize
the system by the ratio $m_1/m_2$.    

In our problem, the use of the atomic units of $A$ and $\bar A$ as
described in Section III brings us significant simplicity. We have
just seen that the reduced mass in atomic units is a simple function
of the constituent mass ratio $m_1/m_2$.  It should also be realized
that the interaction potential and its different components in atomic
units depend only on the various coefficients $F_{A\bar A}$ or
$f_{1,2,3,4}$, which are themselves ratios and are uniquely
characterized by $m_1/m_2$.  Therefore, the Coulomb four-body system
in atomic units is completely characterized by $m_1/m_2$.
Consequently, two different four-body systems with the same $m_1/m_2$
will have the same molecular state eigenenergies and eigenfunctions in
atomic units.  We can infer the stability of a molecule by studying
the change of the molecular eigenenergies as a function of $m_1/m_2$.

\begin{figure} [h]
\hspace*{0.0cm}
\includegraphics[scale=0.45]{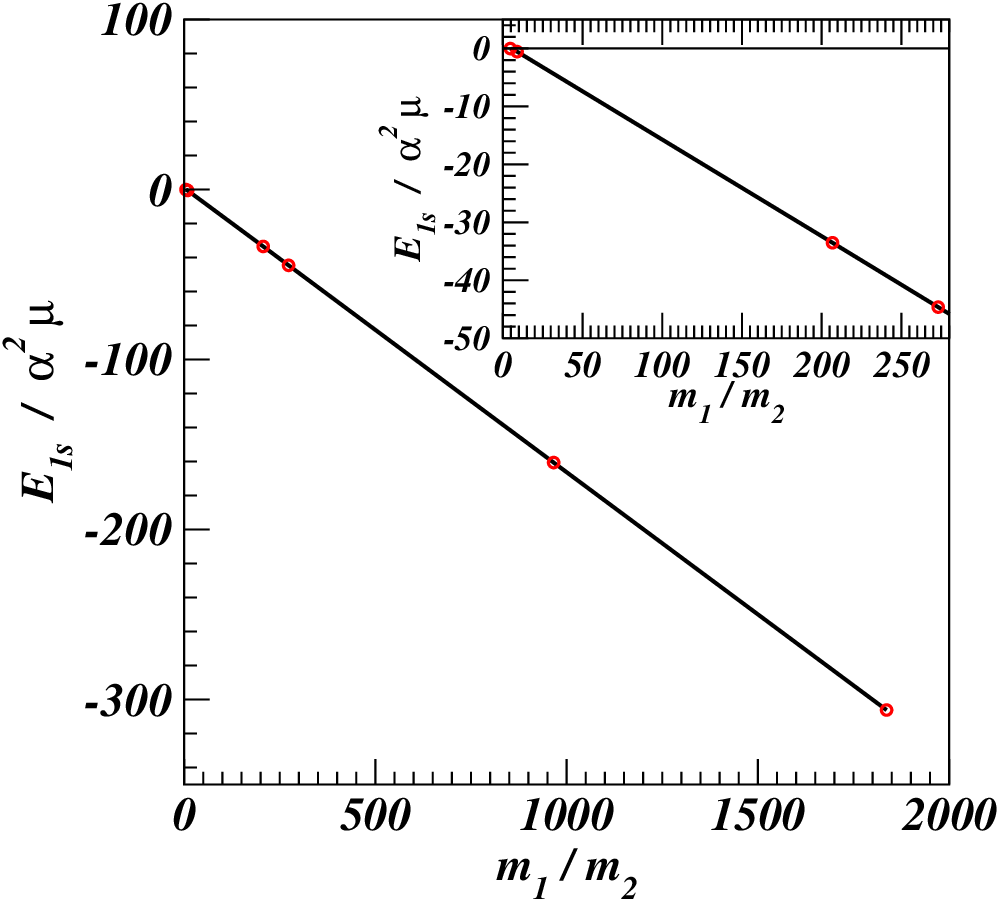}
\caption{ The eigenenergy of the $1s$ molecular state as a function of $m_1/m_2$.
  } \label{fig10}
\end{figure}

We give in Table II the values of $m_1/m_2$, the reduced mass in
atomic units of the $A$ atom, and the molecular state binding
property, for many four-particle systems. The reduced mass decreases
as $m_1/m_2$ decreases.  

\begin{table}[h]
 \caption { Relationship  between $m_1/m_2$, reduced mass,  and the presence of 
bound  molecular states in many four-body systems.}
\vspace*{0.2cm} 
\hspace*{0.0cm}
\begin{tabular}{|c|c|c|c|}
\cline{1-4}
     System    &   &Reduced mass  &  Eigenenergy of  \\
      $A(m_1^+ m_2^-)$-${\bar A}({\bar m}_2^+ {\bar m}_1^-)$ & $m_1/m_2$ &  in atomic units & 1s molecular state  \\ 
    $(m_1> m_2)$  &         &            & \\ \hline 
$(pe^-)$-$(e^+{\bar p})$ & 1836.2 &919.1 & -306.2 \\ \hline
$(K^+ e^-)$-$(e^+K^-)$ & 966.1 &484.1 &-160.6 \\ \hline
$(\pi^+ e^-)$-$(e^+\pi^-)$ &273.1 & 137.6&-44.63 \\ \hline
$(\mu^+e^-)$-$(e^+\mu^-)$ & 206.8 &104.4 & -33.52 \\ \hline
$(p\mu^-)$-$(\mu^+{\bar p})$ & 8.88 &5.50& -0.540\\ \hline
$(K^+\mu^-)$-$(\mu^+K^-)$ & 4.67 &3.44&  -0.0239\\ \hline
$(\tau^+{\bar p})$-$(p\tau^-)$ &1.89 &  2.21& None \\ \hline
\end{tabular}
\end{table}

How does the interaction potential varies as $m_1/m_2$ decreases?  For
the case of $m_1/m_2 \gg 1$, the interaction potential at small $r$ is
dominated by the direct potential over the polarization potential.
The total interaction potential is strongly attractive at small $r$.
As $m_1/m_2$ decreases and approaches $1$, the direct potential is
significantly reduced, and the interaction potential becomes dominated
by the polarization potential.  The net result is a decrease in the
strength of the attractive interaction.

We solve the Schr\"odinger equation (\ref{sch1}) to obtain the
eigenstate energies $\epsilon$ for different molecular systems using
the corresponding reduced masses and interaction potentials.  The
fourth column in Table II indicates whether bound states are present
in various systems.  There are bound states in $(pe^-)$-$(e^+{\bar
  p})$, $(K^+ e^-)$-$(e^+K^-)$, $(\pi^+ e^-)$-$(e^+\pi^-)$,
$(\mu^+e^-)$-$(e^+\mu^-)$, $(p\mu^-)$-$(\mu^+{\bar p})$, and
$(K^+\mu^-)$-$(\mu^+K^-)$.  The results in Table II indicate that
bound molecular states exist for four-particle systems if $m_1/m_2$
is greater than about 4 (or if the reduced mass is greater than or
about 3 atomic units).  The eigenenergies for many four-particle
systems are shown in Fig.\ 9.  We label eigenstates by the angular
momentum and the principal quantum number, which is equal to the
number of nodes plus $l+1$.  The eigenenergies (measured in their
corresponding atomic unit $\alpha^2 \mu_{12}$) move up into the
continuum as $m_1/m_2$ approaches unity and the reduced mass
decreases.

We can examine the molecular state energies of $(p e^-)$-$( e^+ {\bar
  p})$ and compare them with those of a $p\bar p$ atom.  For the $(p
e^-)$-$( e^+ {\bar p})$ molecule, the molecular 1$s$ state energy is
located at -306.2$\alpha^2\mu_{pe}$ and the $ns$ state is higher than
the $np$ state, whereas the atomic $1s$ state energy for a $p\bar p$
atom is -459.5$\alpha^2\mu_{pe}$ and the $ns$ state has the same
eigenenergy as the $np$ state.  The differences in state energies and
ordering arise because in the $(p e^-)$-$( e^+ {\bar p})$ system, the
interaction potential at small $r$ is given by Eq.\ (\ref{dir}) (or
Eq.\ (\ref{35})) that contains the first term of an attractive
screened potential with a screening length $f_1 a/2 \sim a/2$.  This
screening length is so large compared to the $p\bar p$ Bohr radius
that the attractive screened potential is nearly $-\alpha/r$ in
character, as in a $p \bar p$ atom.  However, there is an additional
repulsive screened Coulomb interaction in Eq.\ (\ref{dir}) with a
screening length $f_2 a/2$ that is comparable to the $p\bar p$ Bohr
radius.  This repulsive screened Coulomb interaction arises because in
the $(p e^-)$-$( e^+ {\bar p})$ molecule, the orbiting of the leptons
with respect to the baryons leads to the motion of the baryons.  As a
consequence, the proton and the antiproton have a spatial distribution
and the interaction between the charge distributions of the proton and
the antiproton is reduced from their point-charged values.  This
additional repulsive screened Coulomb interaction in Eq.\ (\ref{dir})
raises the eigenenergy of $(p e^-)$-$( e^+ {\bar p})$ in the 1$s$
state molecular state relative to the eigenenergy of the $p\bar p$
atom in the $1s$ state, and the $ns$ state to lie higher than the $np$
state.

For $(p\mu^-)$-$(\mu^+{\bar p})$ for which $m_1/m_2=8.88$, $1s$ state
energy is at -0.5425 atomic units.  The molecular state is weakly
bound.  For $(K^+\mu^-)$-$(\mu^+K^-)$ for which $m_1/m_2= 4.67$, the
$1s$ state energy lies at -0.0239 atomic units , which is just
barely bound, indicating that $m_1/m_2 \sim 4$ is the boundary between
the region of bound and unbound molecular states.  For $(\tau^+\bar
p)$-$(p\tau^-)$ for which $m_1/m_2=1.89$, there is no bound molecular
state.

We plot in Fig.\ \ref{fig10} the eigenenergy of the molecular $1s$
state as a function of $m_1/m_2$.  As one observes in
Fig.\ \ref{fig10} and Table II, the engenenergy varies approximately
linearly as a function of $m_1/m_2$, with an intercept of $E=0$ at
$m_1/m_2 \sim 4$, indicating that bound molecular states are present
for $m_1/m_2$ greater than about 4.

\section{Annihilation Lifetimes of Molecular States}

Having located the energies of matter-antimatter molecules in various
systems, we would like to calculate their annihilation lifetimes.  The
annihilation probability depends on many factors: the spatial factor
of contact probability that is determined by the particle wave
functions, the type of annihilating particles whether they are leptons
or hadrons, and the spins of the annihilating pair if they are
leptons.  We shall discuss these different factors in turn.

\subsection{Spatial Factor in Particle-Antiparticle Annihilation}

In our matter-antimatter molecules, $m_1$-$m_4$ and $m_2$-$m_3$ are
charge conjugate pairs which can annihilate.  The total annihilation
probability naturally comprises of $P_{14}$ for the annihilation of
$m_1$ and $m_4$, and $P_{23}$ for the annihilation of $m_2$ and
$m_3$. These spatial factors $P_{jk}$ can be obtained by approximating
the constituent wave function to contain only the first term of the
perturbative expansion in Eq.\ (\ref{wf}), as amplitudes of the
excited states relative to the unperturbed states are small and the
excited states have greater spatial extensions that suppress the
annihilation probabilities.

We shall limit our attention to the annihilation of $s$-wave molecular
states, which dominates the annihilation process.  By the term
``annihilation" in the present work, we shall refer to the
annihilation in the $s$-wave molecular states only.  The probabilities
for the annihilation in higher angular momentum states are higher
order in $\alpha$ \cite{Cra06} and involve not only derivatives of the
molecular wave function at the origin but also complicated angular
momentum couplings.  They will need to be postponed to a later date.

We shall calculate the spatial factor $P_{jk}$ for the molecular
($s$-wave) state $\psi ({\bb r})$ built on $A_\nu(m_1 m_2)$ and ${\bar
  A}_{\nu'}(m_3m_4)$ atoms.  It is quantitatively defined as the
probability per unit volume for finding the conjugate $m_j$-$m_k$ pair
to be at the same spatial location, in the molecular ($s$-wave) state
$\psi ({\bb r})$ with $A_\nu(m_1 m_2)$ and ${\bar A}_{\nu'}(m_3m_4)$
atoms.  It is the expectation value of $\delta({\bb r}_{jk})$,
\begin{eqnarray}
P_{jk}&=&\int d{\bb r} d{\bb r}_{12} d{\bb r}_{34}
  A_\nu^* ({\bb r}_{12}) {\bar A}_{\nu'}^*({\bb r}_{34}) 
 \psi^* ({\bb r})\delta({\bb r}_{jk}) \psi ({\bb r})
 A_\nu ({\bb r}_{12}) {\bar A}_{\nu'}({\bb r}_{34}).
~~~~~\end{eqnarray}
We shall consider molecular states built on $\nu=\nu'=0$, then 
\begin{eqnarray}
P_{jk}&=&
 \int d{\bb r} 
|\psi ({\bb r})|^2
 \langle A_0 {\bar A}_0 |
 \delta({\bb r}_{jk}) |A_0  {\bar A}_0\rangle
\nonumber\\
&=&
 \int d{\bb r} 
|\psi ({\bb r})|^2
 \langle A_0 {\bar A}_0 |
 \delta({\bb r}+F_A(jk){\bb r}_{12}+ F_B(jk) {\bb r}_{34} ) |A_0  {\bar A}_0\rangle,~~~
\label{eq31}
\end{eqnarray}
in which $\langle A_0 {\bar A}_0 |\delta({\bb r}_{jk})|A_0 {\bar
  A}_0\rangle$ has the same structure as $V_{\rm dir}$ except that
$v_{jk}({\bb r}_{jk})$ is replaced by a delta function.  For this case
with $\nu=\nu'=0$, the sign of $F_{A,{\bar A}}$ does not matter, and
we can just use $f_{A,{\bar A}}$ for $F_{A,\bar A}$.  Similar to
Eq.\ (\ref{32}), we have
\begin{eqnarray}
&&\langle A_{0} {\bar A}_{0}| \delta(\bbox{r}_{jk}) | A_{0} {\bar A}_{0}\rangle
= \int \frac{d\bbox{p}}{(2\pi)^3} e^{i \bbox{p}\cdot
\bbox{r}}
\frac {16 } {((pf_A a_0)^2+4)^2}   \frac {16 } {((pf_{\bar A} a_0)^2+4)^2}.  ~~~~~
 \end{eqnarray}
For both $\{jk\}=\{14\}$ and $\{23\}$, $f_{\bar A}(jk)=f_A(jk)$, and 
we get
\begin{eqnarray}
 D_{jk}({\bb r})&\equiv& \langle A_{0} {\bar A}_{0}| \delta_{jk}(\bbox{r}_{jk}) | A_{0} {\bar A}_{0}\rangle
\nonumber\\
&=&
\frac{e^{-\lambda_A(jk) r}}{4\pi}
\frac{[\lambda_A(jk)]^3}{48}
\left \{ [ \lambda_A(jk) r]^2  
+ 3\lambda_A(jk) r+3\right \},
\label{eq37}
\end{eqnarray}
where
\begin{eqnarray}
\lambda_A(jk)&=&\frac{2}{f_A(jk) a},\nonumber \\
f_A(14)&=&\frac{m_2}{m_1+m_2},~~~~~~
f_A(23)=\frac{m_1}{m_1+m_2}.\nonumber
\end{eqnarray}
 Eq.\ (\ref{eq31}) becomes
\begin{eqnarray}
P_{jk}=\int d{\bb r} 
|\psi ({\bb r})|^2
D_{jk}({\bb r}).
\label{55}
\end{eqnarray}
As the wave function $\psi({\bb r})$ has been obtained from our
solution of the Schr\" odinger equation, the spatial factors can be
calculated numerically.

Following Wheeler \cite{Whe46}, we can consider an antiparticle $\bar
m$ to be at rest while its conjugate particle $m$ sweeps by with a
annihilation cross section $\sigma_{\rm ann}^{m\bar m}$ at a velocity
$v$, clearing a volume $\sigma_{\rm ann}^{m\bar m}v$ per unit time.
The probability of finding the pair of particles $j$ and $k$ to be in
contact, per unit spatial volume, is $P_{jk}$.  As a consequence, the
number of particle-antiparticle contacts per unit time, which is equal
to the rate of annihilation $w_{jk}^{m \bar m}$, is given by
\begin{eqnarray}
w_{jk}^{m \bar m}&=& \sigma_{\rm ann}^{m \bar m} v  \times P_{j  k}.
\end{eqnarray}

\subsection{Annihilation of Lepton Constituents}

The matter-antimatter molecules we have been considering consist of
both leptons and hadrons.  The annihilation cross section $\sigma_{\rm
  ann}^{m \bar m}$ depends on the particle type and the total spin of
the annihilating pair.  We shall discuss the annihilation of lepton
pairs in this subsection. The annihilation of hadron pairs will be
discussed in the next subsection.

The mechanism for the annihilation of a lepton pair is well known
\cite{Whe46,Ber82}.  It proceeds through the electromagnetic
interaction with the fine-structure coupling constant $\alpha$ and the
emission of two or three photons.  A lepton pair with spin $S$=$0$ can
annihilate only into two photons, and a lepton pair with spin $S$=$1$
can annihilate only into three photons \cite{Whe46}.  The lepton pair
annihilation cross section multiplied by velocity $v$ is given by
\cite{Whe46,Ber82}
\begin{eqnarray}
&&\sigma_{S_{jk}=0}^{l \bar l}v ~~({\rm for~annihilation~into~2~photons})= \pi \left  (\frac{\alpha}{ \mu c^2}\right )^2 
\label{56}
\\
&&\sigma_{S_{jk}=1}^{l \bar l}v~~ ({\rm for~annihilation~into~3~photons})=
\frac{4(\pi^2-9)}{9} \alpha
\left (\frac{\alpha}{ \mu c^2}\right )^2 
\end{eqnarray} 
As a consequence, 
the rates of
annihilation of a lepton pair in the singlet and triplet states are 
\cite{Whe46,Ber82}
\begin{eqnarray}
w_{S_{jk}=0}^{l \bar l}({\rm annihilation~into~2~photons})&=& \pi \left  (\frac{\alpha}{ \mu c^2}\right )^2 P_{jk}
\label{s0}
\\
w_{S_{jk}=1}^{l \bar l}({\rm annihilation~into~3~photons})&=&
\frac{4(\pi^2-9)}{9} \alpha
\left (\frac{\alpha}{ \mu c^2}\right )^2 P_{jk}~~~~~~~~~~
\label{s1}
\end{eqnarray} 
Thus, to obtain the rate of annihilation of a lepton-antilepton pair
$\{jk\}$ in a molecular state, it is necessary to find out the
probabilities for the pair to be in different $S_{jk}$ pair spin
states in the molecule.

Our molecular states have been constructed by building them with
$A(12)$ and ${\bar A}(34)$ atoms in their ground states.  As we
restrict our considerations to only $s$-wave molecular states, there
is no orbital angular momentum between the $A$ and $\bar A$ atoms.  In
picking the lepton $l$ from atom $A$ and the antilepton $\bar l$ from
the other atom $\bar A$, the probabilities for different
lepton-antilepton pair spin states depend on the angular momentum
coupling of the lepton-antilepton pair with the remaining
constituents.

We consider first the case when the remaining constituents are also
fermions. The four constituents can be labeled as
$\{m_1$=$f,m_2$=$l,m_3$=${\bar l},m_4$=${\bar f}\}$.  Each of the
$A(12)$ and ${\bar A}(34)$ atoms has an atomic spin
$S_{12}^A,S_{34}^{\bar A}=0$ or 1.  As a consequence, the $s$-wave
molecular state has a total molecular spin $S_{\rm tot}^{A\bar
  A}=0,1,$ or 2.  By the definition of the 9-$j$ symbols \cite{Des63},
the probability amplitude for the occurrence of fermion-antifermion
spin states of $S_{14}^{f \bar f}$ and $S_{32}^{l \bar l}$ in a state
with atom spins $S_{12}^A$ and $S_{34}^{\bar A}$ and molecular spin
$S_{\rm tot}^{A\bar A}$ is
\begin{eqnarray}
&&\langle S_{14} ^{f \bar f} S_{23}^{l \bar l}; S_{\rm tot}^{A\bar A} | S_{12}^A S_{43} ^{\bar A}; S_{\rm tot}^{A\bar A}\rangle
\nonumber\\
&&=\sqrt{(2S_{14}^{f \bar f}+1) (2S_{23}^{l \bar l}+1)
                 (2S_{12}^A+1)(2S_{43}^{\bar A}+1)}
\left \{ 
\begin{matrix}
s_1       & s_2       & S_{12} ^A\cr
s_4       & s_3       & S_{43}^{\bar A} \cr
S_{14} ^{f \bar f}& S_{23}^{l \bar l} & S_{\rm tot}^{A\bar A}
\end{matrix}
\right \},~~~~
\end{eqnarray}
where $s_1$=$s_2$=$s_3$=$s_4$=1/2.  The probability $|\langle
S_{14}^{f \bar f} S_{23}^{l \bar l}; S_{\rm tot}^{A\bar A} | S_{12}^A
S_{43}^{\bar A} ; S_{\rm tot}^{A\bar A}\rangle|^2$ for different
atomic spins of $S_{12}^A$ and $S_{43}^{\bar A}$ combining into
different $S_{14}^{f \bar f}$ and $S_{23}^{l \bar l}$, for a fixed
total molecular spin $S_{\rm tot}$, are given in Table III.
\begin{table}[h]
  \caption { Probabilities for different atomic spins of $S_{12}^A$
    and $S_{34}^{\bar A}$, combining into $S_{14}^{f \bar f}$ and
    $S_{23}^{l \bar l}$, for a fixed total molecular spin $S_{\rm
      tot}^{A\bar A}$.}
\vspace*{0.2cm} 
\hspace*{0.0cm}
\begin{tabular}{|c|c|c|c|c|c|}
       \cline{2-6} 
  \multicolumn{1}{c|}{}  & $S_{12}^A$ & $S_{34}^{\bar A}$ & $S_{14}^{f \bar f}$  & $S_{23}^{l \bar l}$ &  probability     \\ \hline 
$S_{\rm tot}^{_{A\bar A}}=2$ & 1  & 1 & 1 & 1 & 1 \\ \hline
$S_{\rm tot}^{_{A\bar A}}=1$ & 1  & 1 & 1 & 0 & 1/2 \\ 
                           & 1  & 1 & 0 & 1 & 1/2 \\ \hline
$S_{\rm tot}^{_{A\bar A}}=1$ & 1  & 0 & 1 & 1 & 1/2 \\ 
                           & 1  & 0 & 1 & 0 & 1/4 \\ 
                           & 1  & 0 & 0 & 1 & 1/4 \\ \hline
$S_{\rm tot}^{_{A\bar A}}=1$ & 0  & 1 & 1 & 1 & 1/2 \\ 
                            & 0  & 1 & 1 & 0 & 1/4 \\ 
                            & 0  & 1 & 0 & 1 & 1/4 \\ \hline
$S_{\rm tot} ^{_{A\bar A}}=0$ & 1  & 1 & 1 & 1 & 1/4 \\ 
                           & 1  & 1 & 0 & 0 & 3/4 \\ \hline
$S_{\rm tot}^{_{A\bar A}}=0$ & 0  & 0 & 1 & 1 & 3/4 \\ 
                           & 0  & 0 & 0 & 0 & 1/4 \\ 
 \hline
\end{tabular}
\end{table}

\begin{table}[h]
  \caption { The $s$-wave $(\mu^+ e^-)$-$(e^+\mu^-)$ molecular state
    energies and annihilation lifetimes in different $S_{\rm
      tot}^{A\bar A}$, $S_{12}^A$, and $S_{34}^{\bar A}$ spin
    configurations.}
\vspace*{0.2cm} 
\hspace*{0.0cm}
\begin{tabular}{|c|c|c|c|c|c|}
       \cline{1-6} 
  $(\mu^+ e^-)$-$(e^+\mu^-)$  &Energy($\alpha^2 \mu$)  &     &    &   & Annihilation \\
 Molecular state   &  & $S_{\rm tot}^{_{_{A\bar A}}}$ & $S_{12}^A$ & $S_{34}^{_{\bar A}}$ & Lifetime $\tau_{\rm ann}$(sec)  \\ \hline 
 1$s$& -33.5& 2 & 1 & 1 & 0.861$\times 10^{-8}$ \\
       &                & 1 & 1 & 1 & 0.154$\times 10^{-10}$ \\
       &                & 1 & 1 & 0 & 0.308$\times 10^{-10}$ \\
       &                & 1 & 0 & 1 & 0.308$\times 10^{-10}$ \\
       &                & 0 & 1 & 1 & 0.103$\times 10^{-10}$ \\
       &                & 0 & 0 & 0 & 0.308$\times 10^{-10}$ \\ \hline
 2$s$& -9.30 & 2 & 1 & 1 & 0.405$\times 10^{-7}$ \\
       &                & 1 & 1 & 1 & 0.726$\times 10^{-10}$ \\
       &                & 1 & 1 & 0 & 0.145$\times 10^{-9}$ \\
       &                & 1 & 0 & 1 & 0.145$\times 10^{-9}$ \\
       &                & 0 & 1 & 1 & 0.484$\times 10^{-10}$ \\
       &                & 0 & 0 & 0 & 0.145$\times 10^{-9}$ \\
\hline 
\end{tabular}
\end{table}

By taking into account spin probabilities, the total rate of
annihilation for matter-antimatter molecules consisting of 4 fermions
with total molecular spin $S_{\rm tot}^{A\bar A}$, initial atomic
spins $S_{12}^A$ and $S_{34}^{\bar A}$ is
\begin{eqnarray}
w_{\rm ann}(S_{12}^A S_{34}^{\bar A}S_{\rm tot}^{A\bar A})=
\sum_{S_{14} S_{23}} 
|\langle S_{14} S_{23}; S_{\rm tot}^{A\bar A}) | S_{12} ^A S_{43}^{\bar A} ; S_{\rm tot}^{A\bar A}\rangle|^2
[w_{_{S_{14}}}^{f \bar f}+w_{_{S_{23}}}^{l \bar l}],
\label{62}
\end{eqnarray}
where for lepton annihilations $w_{_{S_{23}}}^{l \bar l}$ is given by
Eqs.\ (\ref{s0}) and (\ref{s1}).  The annihilation lifetime is then
given by
\begin{eqnarray}
\tau_{\rm ann}(S_{12}^A S_{34}^{\bar A}S_{\rm tot}^{A\bar A})=\frac{1}{w_{\rm ann}(S_{12}^A S_{34}^{\bar A}S_{\rm tot}^{A\bar A})}
\end{eqnarray}
The rate of annihilation of the $(\mu^+ e^-)$-$(\mu^+e^+)$ molecule in
its $ns$ states can be calculated from Eq.\ (\ref{62}) by identifying
$f$=$\mu$ and $l$=$e$.  The results for the annihilation lifetimes are
shown in Table IV where we list the $(\mu^+ e^-)$-$(e^+\mu^-)$
molecular state energies, the molecular spins $S_{\rm tot}^{A\bar A}$,
the atomic spins $S_{12}^A$, $S_{34}^{\bar A}$ and the annihilation
lifetimes $\tau_{\rm ann}$.  In calculating the annihilation lifetimes
in their physical units, we have used Table I to convert atomic units
to physical units.

The $(\mu^+ e^-)$-$(e^+\mu^-)$ molecular states with spin $S_{\rm
  tot}^{A\bar A}$=2 have the longest annihilation lifetimes,
corresponding to lepton-antilepton pairs in their spin triplet states.
The molecular 1$s$ and $2s$ $S_{\rm tot}^{A\bar A}=2$ states have
annihilation lifetimes of $0.861\times 10^{-8}$ and $0.405\times
10^{-7}$ sec, respectively.  Their relatively long annihilation
lifetimes may make them accessible for experimental observations.

\subsection{Annihilation of Hadron Constituents}

Hadrons are composite particles consisting of quarks and/or antiquarks
whose quantized masses are governed by non-perturbative quantum
chromodynamics.  As a consequence, there are no simple selection rules
for hadron annihilation similar to those for the annihilation of
leptons.

The lightest quantized hadrons are pions.  Because of the difference
in their masses, the annihilation of heavier hadron pairs such as
$p\bar p$ and $K^+K^-$ differ from the annihilation of $\pi^+\pi^-$.
The annihilations of $p\bar p$ and $K^+K^-$ pairs proceed through
strong interactions as many pairs of pions and other hadrons can be
produced.  In contrast, a $\pi^+\pi^-$ pair can annihilate through 
strong interactions only when its center-of-mass energy $\sqrt{s}$
exceeds the threshold of $4m_\pi$, with the production of an
additional pion pair.  For pions in a matter-antimatter molecule, the
momentum of the pion is of order $\alpha m_\pi$ and it has energy much
below the strong-interaction annihilation threshold.  We can infer
that pions in a matter-antimatter molecule annihilate predominantly
through the electromagnetic interaction with the emission of photons
or dileptons.
\begin{figure} [h]
\hspace*{0.0cm}
\includegraphics[scale=0.45]{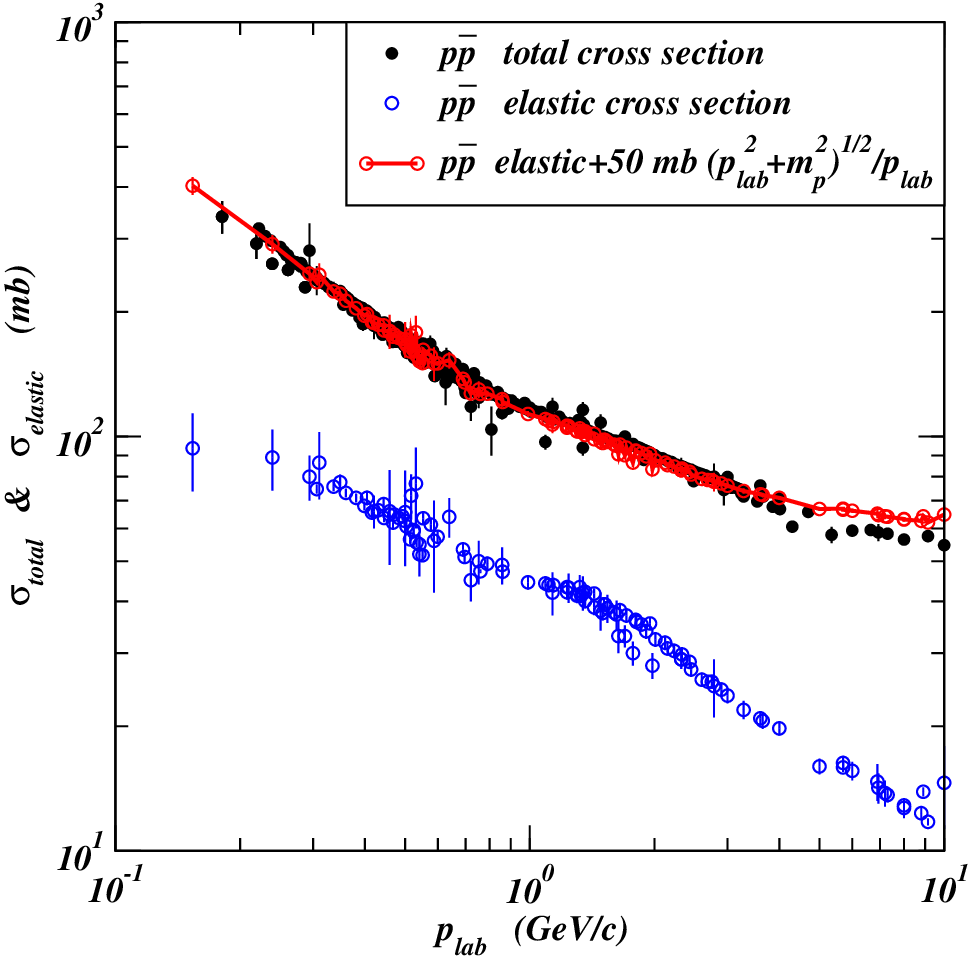}
\caption{ $p\bar p$ total cross section in mb as a function of $p_{\rm lab}$}
\label{fig11}
\end{figure}

We shall first discuss the annihilation of heavy hadrons such as
$p\bar p$ and $K^+K^-$.  We envisage that these hadrons annihilate
essentially through a geometrical consideration depicting the
occurrence of annihilations within a geometrical area $\sigma_0$, with
important initial state interactions that lead to a Gamow-factor $1/v$
type enhancement at low energies \cite{Cha95,Won97}.  We can therefore
parametrized the annihilation cross section as
\begin{eqnarray}
\sigma_{\rm ann}=\frac{\sigma_0}{v},
\end{eqnarray}
where $v$ is the relative velocity between the colliding hadrons.
From the PDG data \cite{PDG10}, 
the total and elastic $p\bar p$  cross sections in Fig.\ \ref{fig11}
obey the following relationship:
\begin{eqnarray}
\sigma_{\rm total}^{p\bar p} =\sigma_{\rm elast}^{p\bar p}+ \frac{ 50 {\rm mb}}{v},
\end{eqnarray}
where
\begin{eqnarray}
v=\frac{p_{\rm lab}}
{\sqrt{p_{\rm lab}^2+m_p^2}}. 
\end{eqnarray}•
As $p\bar p$ inelastic cross section is the same as 
the $p\bar p$ annihilation cross section, the experimental data gives $\sigma_0=50$ mb, and 
\begin{eqnarray}
\sigma_{\rm ann}^{p\bar p} v = \sigma_0=50 {\rm ~mb}.
\end{eqnarray}
For other systems, we can use the additive quark model \cite{Won94,Won94a}
to infer that
\begin{eqnarray}
\sigma_{\rm ann}^{h \bar h} \propto n_q^h \times n_q^{\bar h} , 
\end{eqnarray}
where $n_q^h$ and $n_q^{\bar h}$
are  the number of quarks in $h$ and $\bar h$.
We get 
\begin{eqnarray}
\sigma_{\rm ann}^{h\bar h} v \sim \frac{ n_q^h n_q^{\bar h}}{9}{50 {\rm mb}}. 
\label{63}
\end{eqnarray}
Consequently, the rate of annihilation of constituents 1 and 4 through
strong interactions per unit time is
\begin{eqnarray}
w_{14}^{h \bar h}&=&
 \frac{ n_q^h n_q^{\bar h}}{9} 
50 {\rm mb}\times  P_{14} ,~~~\{h\bar h\}=\{p\bar p\} {\rm ~or~} \{K^+K^-\}.
\end{eqnarray}•
In the $(pe^-)$-$(e^+{\bar p})$ molecule, there are two leptons and two
baryons.  The spins of the four fermions are coupled together and we
have
\begin{eqnarray}
w_{\rm ann} (S_{12}^AS_{34}^{\bar A}S_{\rm tot}^{A\bar A})=w_{14}^{p\bar p}+
\sum_{S_{14}^{p\bar p} S_{23}^{e^+e^-}} 
|\langle S_{14}^{p\bar p} S_{23}^{e^+e^-}; S_{\rm tot}^{A\bar A} | 
         S_{12}^A S_{43}^{\bar A} ; S_{\rm tot}^{A\bar A} \rangle|^2
w_{_{S_{23}}}^{e^-e^+}.
\end{eqnarray}
In the $(K^+ e^-)$-$(e^+K^-)$ molecule, there is a lepton pair and a
$K$-$\bar K$ pair.  As the kaons have spin zero, the molecular spin
$S_{\rm tot}$ comes only from the leptons and we have
\begin{eqnarray}
w_{\rm ann}(S_{\rm tot})=w_{14}^{K^+K^-}+
w_{S_{\rm tot}}^{e^+ e^-} .
\end{eqnarray}
In practice, for the molecular $ns$ states we have considered, the
rate of hadron annihilation is much greater than the rate of lepton
annihilation, if the hadrons can annihilate through strong
interactions.  In this case, because of the dominance of the hadron
annihilation through strong interactions, the total annihilation rate
is essentially independent of the spin of the molecule $S_{\rm tot}$.

For the case involving $\pi^+$ and $\pi^-$ constituents, annihilation
cannot proceed through strong interactions as the pion energies are
below the inelastic threshold.  The pion pair can annihilate into
photons and dileptons.  Because the $\pi^+$-$\pi^-$ system has the
same total angular momentum and parity quantum numbers as the spin
singlet state of a lepton pair, and the Feynman diagrams for the
emission of two photons in QED for $l+\bar l \to 2 \gamma$ and $\pi^+
+\pi^- \to 2 \gamma$ have the same structure, we can approximate the
$\pi^+$+$\pi^-$$\to 2 \gamma$ cross section to be the same form as the
spin-singlet $l+{\bar l} \to 2 \gamma$ cross section in
Eq.\ (\ref{56}),
\begin{eqnarray}
\sigma_{\rm ann}^{\pi^+\pi^-} v_{\pi^+ \pi^-} \sim \pi \left  (\frac{\alpha}{ \mu_{\pi \pi} c^2}\right )^2.
\end{eqnarray}
The cross section for pion annihilation into dileptons [as given by
  Eq. (14.33) of Ref.\ \cite{Won94}] is of order $\alpha^3$ for pions
with momentum $p\sim \alpha m_\pi$.  We can neglect the dilepton
contribution from $\pi^+$+$\pi^-$$\to l$+$\bar l$ in the present
estimate.  Consequently, the rate of annihilating pion constituents 1
and 4 by electromagnetic interactions per unit time is
\begin{eqnarray}
w_{14}^{\pi^+ \pi^-}&=& \pi \left (\frac{\alpha}{ \mu_{\pi \pi}
  c^2}\right )^2 \times P_{14} .
\end{eqnarray}•
In the $(\pi^+ e^-)$-$(e^+\pi^-)$ molecule, there is a lepton pair and
a $\pi^+$-$\pi^-$ pair.  As the pions have spin zero, the molecular
spin comes only from the leptons.  For the $(\pi^+ e^-)$-$(e^+\pi^-)$
molecule, we have
\begin{eqnarray}
w_{\rm ann}(S_{\rm tot})=w_{14}^{\pi^+ \pi^-}+
w_{S_{\rm tot}}^{e^+ e^-}.
\end{eqnarray}
Because the annihilation of both the lepton pair and the pion pairs
are electromagnetic in origin, they are comparable in magnitude.  The
annihilation rate depends on the spin of the molecule $S_{\rm tot}$.

\begin{table}[h]
  \caption { Molecular $1s$ state energies and annihilation lifetimes
    for $(pe^-)$-$(e^+\bar p)$, $(p\mu^-)$-$(\mu^+\bar p)$,
    $(K^+e^-)$-$(e^+K^-)$, $(K^+\mu^-)$-$(\mu^+K^-)$ and
    $(\pi^+e^-)$-$(e^+\pi^-)$.}
\vspace*{0.2cm} 
\hspace*{1.5cm}
\begin{tabular}{|c|c|c|c|}
       \cline{1-4} 
 System   & State & Energy($\alpha^2 \mu$) & Annihilation  \\ 
   &  &  & Lifetime(sec)  \\ \hline 
$(pe^-)$-$(e^+\bar p)$ & 1$s$ &
 -306.208  &  0.549$\times 10^{-17}$ \\
                                       &2$s$ &
  -92.565	&  0.347 $\times 10^{-17}$ \\ 
                                       &3$s$ &
 -43.255	&	  0.107$\times 10^{-15}$ \\
                                       &4$s$ &
 -24.595	&	  0.244$\times 10^{-15}$  \\
                                       &5$s$ &
 -15.595	&	  0.464$\times 10^{-15}$  \\
                                       &6$s$ &
   -4.094	&	  0.259$\times 10^{-14}$  \\
\hline  
$(p\mu^-)$-$(\mu^+{\bar p})$ & 1$s$ &
   -0.540   	&	  0.532$\times 10^{-17}$  \\
\hline
$(K^+e^-)$-$(e^+K^-)$ & 1s &
 -160.601	&	  0.853$\times 10^{-16}$  \\
                                         & 2$s$ &
   -48.108	&	  0.526$\times 10^{-15}$  \\
                                         & 3$s$ &
  -22.141	&	  0.166$\times 10^{-14}$  \\
                                         & 4$s$ &
  -12.316	&	  0.372$\times 10^{-14}$  \\
                                         & 5$s$ &
   -7.581	&	  0.715$\times 10^{-14}$  \\
                                         & 6$s$ &
   -4.947	&	  0.122$\times 10^{-13}$  \\
\hline  
$(K^+\mu^-)$-$(\mu^+K^-)$ &  1$s$ &
   -0.0239   	&	  0.154$\times 10^{-15}$  \\
\hline
$(\pi^+e^-)$-$(e^+\pi^-)$& 1$s$ &
 -44.625     &($S_{\rm tot}$=0) 0.592$\times 10^{-11}$ \\
     &  &      &($S_{\rm tot}$=1) 0.622$\times 10^{-11}$  \\
                                           & 2$s$ &
 -12.698     &($S_{\rm tot}$=0) 0.298$\times 10^{-10}$  \\
     &  &      &($S_{\rm tot}$=1) 0.392$\times 10^{-10}$  \\
                                           & 3$s$ &
  -5.341      &($S_{\rm tot}$=0) 0.616$\times 10^{-10}$  \\
     &  &      &($S_{\rm tot}$=1) 0.121$\times 10^{-9}~$  \\
\hline 
\end{tabular}
\end{table}

In Table V, we list the $l$=$0$ molecular states, their energies, and
their lifetime for $(pe^-)$-$(e^+\bar p)$, $(p\mu^-)$-$(\mu^+\bar p)$,
$(K^+e^-)$-$(e^+K^-)$, $(K^+\mu^-)$-$(\mu^+K^-)$, and
$(\pi^+e^-)$-$(e^+\pi^-)$.  We observe that the annihilation lifetimes
are short for the $(pe^+)$-$(e^+\bar p)$ states, of the order of
$10^{-15}$-$10^{-18}$ sec, increasing to order
$10^{-10}$-$10^{-11}$ sec for $(\pi^+e^-)$-$(e^+\pi^-)$ states.

\section{Summary and Discussions}

To examine the stability of matter-antimatter molecules with
constituents $(m_1^+ m_2^- {\bar m}_2^+ {\bar m}_1^-)$ and
$m_1$$>$$m_2$, we reduce the four-body problem into a simpler two-body
problem.  This is achieved by breaking up the four-body Hamiltonian
into the unperturbed Hamiltonians of two atoms $A(m_1^+ m_2^-)$ and
${\bar A}( {\bar m}_2^+ {\bar m}_1^-)$, plus residual interactions and
the kinetic energies of the atoms.  The unperturbed Hamiltonians of
the two atoms can be solved exactly.  Molecular states can be
constructed by using the atoms $A$ and $\bar A$ as simple building
blocks.  The interaction potential $V({\bb r})$ between the atoms is
then the sum of the direct potential $V_{\rm dir}({\bb r})$ arising
from the interaction between the constituents and the polarization
potential $V_{\rm pol}({\bb r})$ arising from the virtual excitation
of the atomic states in the presence of the other atom.  The
eigenenergies of the molecular states can be obtained by quantizing
the four-particle Hamiltonian \cite{Won04}.

The Coulomb four-body system in atomic units of $A$ and $\bar A$ atoms
is completely characterized by $m_1/m_2$.  Consequently, two different
four-body systems with the same $m_1/m_2$ will have the same molecular
state eigenenergies and eigenfunctions in atomic units.  We can infer
the stability of a molecule by studying the change of the molecular
eigenenergies as a function of $m_1/m_2$.

The effective reduced mass of  $A(m_1^+ m_2^- )$-${\bar
  A}({\bar m}_2^+ {\bar m}_1)$ for molecular motion
is large when $m_1/m_2\gg 1 $ and decreases as
$m_1/m_2$ approaches unity.  The relative importance of the direct and
polarization potential also changes with $m_1/m_2$.  For a
matter-antimatter molecule with $m_1/m_2 \gg 1$, we find that the
direct potential dominates in regions of $r < a$ and gives rise to
deeply bound molecular states.  As $m_1/m_2$ approaches unity, the
magnitude of $V_{\rm dir}$ decreases and the polarization potential by
itself is too weak to hold a bound state.  As a consequence, the state
energies (in atomic units) rises and comes up to the continuum as
$m_1/m_2$ approaches the unity limit.

We find that matter-antimatter molecules possess bound states if
$m_1/m_2$ is greater than about 4.  This stability condition suggests
that the binding of matter-antimatter molecules is a rather common
phenomenon.  This molecular stability condition is satisfied, and many
bound molecular states of different quantum numbers are found, in many
four-body systems: $(\mu^+e^-)$-$(e^+\mu^-)$,
$(\pi^+e^-)$-$(e^+\pi^-)$, $(K^+e^-)$-$(e^+K^-)$, $(pe^-)$-$(e^+\bar
p)$, $(p\mu^-)$-$(\mu^+{\bar p})$, and $(K^+\mu^-)$-$(\mu^+K^-)$.
Bound molecular states are not found in $(\tau^+{\bar
  p})$-$(p\tau^-)$ which has $m_1/m_2=1.89$.

When one applies the stability condition to the $(e^+e^-)$-$(e^-e^+)$
system, one may naively infer at first that the $(e^+e^-)$-$(e^-e^+)$
system will not hold a bound state.  On the other hand, bound
$(e^+e^-)^2$ molecular state has been experimentally observed
\cite{Cas07}, and the binding energy has been calculated theoretically
to be 0.016 a.u.\ relative to two $(e^+e^-)$ atoms \cite{Koz93}.  It
should however be realized that the stability condition we have
obtained applies to four-body systems with distinguishable
constituents without identical particles.  For systems with identical
particles such as the $(e^+e^-)^2$ system, it is necessary to take
into account the antisymmetry of the many body wave function with
respect to the exchange of the pair of identical particles.  Depending
on the spin symmetry of the identical particle pair in question, the
symmetry or antisymmetry with respect to the spatial exchange of the
pair will lead to a lowering or raising of the energy of the state.
The identical particles in the bound $(e^+e^-)^2$ system leading to
the bound state have been selected to be spatially symmetric states
\cite{Whe46,Koz93}, which corresponds to spin-antisymmetric with
respect to the exchange of the pair of identical particles
\cite{Koz93}.  As a consequence, their spatial symmetry lowers the
state energy relative to the state energy when there is no such a
symmetry.  The small value of the theoretical binding energy (0.016
a.u.\ \cite{Koz93}) suggests the occurrence of such a lowering of the
energy from the unbound to the bound energy region.  In order to
confirm this suggestion, it will be of interest in future work to
extend the present formulation to include the case with identical
particles and their exchange symmetries.  How the additional symmetry
considerations may modify the stability of those molecules with
identical particles is worthy of future investigations.  The states we
have obtained may not contain all the correlations.  Additional
correlations may be included by adding correlation factors and using
the solution of the present investigation as doorway states.  Future
addition of correlations superimposing on the states we have obtained
will be of interest.

We can divide the annihilation lifetimes of matter-antimatter
molecules into different groups that correlate with the types of
constituents.  Those molecules constructed from different leptons have
the longest annihilation lifetimes, of the order of
$10^{-8}$-$10^{-11}$ sec, depending on the spins of the molecular
state.  The second group involves leptons and pions in which the pions
cannot annihilate through strong interactions and can annihilate only
through the electromagnetic interactions.  The annihilation lifetimes
are of order $10^{-10}$-$10^{-11}$ sec.  Molecular states containing
kaons have annihilation lifetimes of order $10^{-14}$-$10^{-16}$ sec
while those containing proton and antiproton $10^{-15}$-$10^{-18}$
sec.  The relatively long annihilation lifetimes for leptonic
$(\mu^+e^-)$-$(e^+\mu^-)$ molecules may make them accessible for
experimental detection.

We have examined only molecular states constructed from the $A\bar A$
family in which the building-block atoms $A_0$ and ${\bar A}_0$ are in
their ground states.  We can likewise construct in future work states
in the $A\bar A$ family in which $A$ and $\bar A$ are in their excited
states.  These states will have different interatomic interactions,
obey different stability conditions, and have different properties
with regard to annihilation and production. In another future
direction of extension, we can also construct molecular states based
on the $M\bar M$ family, using states of $M(m_1^+ \bar m_1^-)$ and
${\bar M}({\bar m}_2^+ m_2^-)$ as building blocks.  The $A \bar A$
molecular states and the $M \bar M$ molecular states have different
topological structures and properties.  Molecular states built on
different branches of the family tree will likely retain some of their
family characteristics.  These possibilities bring into focus the rich
degrees of freedom and the vast varieties of states that need to be
sorted out in the Coulomb four-body problem associated with
matter-antimatter molecules.
 
In addition to the problem of molecular states as a structure problem
investigated here, future theoretical and experimental studies should
also be directed to the question of production and detection methods
from reaction points of view.  While the observation of new
matter-antimatter molecules may be a difficult task, the prospect of
classifying the $(m_1^+ m_2^- {\bar m_2}^- {\bar m}_2^+)$ system into
the genealogy of families and basic building blocks, if it is at all
possible, will bring us to a better understanding of the complexity of
the spectrum that is associated with the complicated four-body problem
opened up by the pioneering work of Wheeler \cite{Whe46}.

\appendix

\section{Evaluation of the Matrix Element\\$ \langle A_\lambda {\bar A}_{\lambda'}| v(\bbox{r}_{jk}) |A_0 {\bar A}_{0}\rangle$}

The evaluation of the excitation transition matrix element requires
first the Fourier transform of the transition density.  The transition
density for the excitation from the ground $|100\rangle$ state to the
excited $|n1m\rangle$ state is
\begin{eqnarray}
\rho_{n1m,100} ({\bf r})&=&\psi_{n1m}^*({\bb r})\psi_{100}({\bb r}) 
\nonumber\\
&=&\frac{N_{10}N_{n1}}{\sqrt{4\pi}}\exp\left \{ - \frac{2r(n+1)}{2na_0} \right \}
\left(\frac{2r}{na_0}\right) L_{n-2}^{3}(\frac{2r}{na_0})Y_{1m}^*(\theta,\phi).
~~~~~~\label{eq55}
\end{eqnarray}
Making the scale transformation $r=r'/(n+1)=r'\beta$, we have 
\begin{eqnarray}
\rho_{n1m,100} ({\bf r})
=\frac{N_{10}N_{n1}}{\sqrt{4\pi}}\exp\left \{ - \frac{2r'}{2na_0} \right \}
\left(\frac{2r'}{na_0}\right) \beta L_{n-2}^{3}\left (\frac{2r'}{na_0}\beta\right )Y_{1m}^*(\theta,\phi).
\end{eqnarray}
We can expand $L_{j}^{\alpha}(\beta x)$ as a sum over Laguerre
polynomials $L_{j-\kappa}^\alpha(x)$ as given in Eq.\ (22.12.6), page
785 in \cite{Abr70},
\begin{eqnarray}
L_j^\alpha(\beta x)=\sum_{\kappa=0}^j 
\left ( \begin{matrix}
j+\alpha\cr
  \kappa 
\end{matrix} \right )
\beta^{j-\kappa} (1-\beta)^\kappa L_{j-\kappa}^\alpha (x).
\end{eqnarray}
The transition density becomes
\begin{eqnarray}
\label{75}
\rho_{n1m,100} ({\bf r})
=\frac{N_{10}N_{n1}\beta}{\sqrt{4\pi}}
 \sum_{\kappa=0}^{n-2} \left ( \begin{matrix}
n-2+3\cr
  \kappa 
\end{matrix} \right )\beta^{n-2-\kappa} (1-\beta)^\kappa
A_{n1m,100}^\kappa ({\bf r}),
\end{eqnarray}
where
\begin{eqnarray}
A_{n1m,100}^\kappa ({\bf r})
=
\left [ \exp\left \{ - \frac{2r'}{2na_0} \right \}
 \left(\frac{2r'}{na_0}\right) L_{n-2-\kappa}^{3}
\left (\frac{2r'}{na_0}\right )Y_{1m}^*(\theta,\phi)\right ]. \nonumber
\label{eq57}
\end{eqnarray}
Equation (\ref{75}) is a sum of many terms, each of which is a
hydrogen wave function.  The Fourier transform of $A_{n1m,100}^\kappa
({\bf r})$ is given by
\begin{eqnarray}
{\tilde A}_{n1m,100}^\kappa({\bf p})&=&\int d{\bf r}~~e^{-i{\bf p}\cdot {\bf r}} A_{n1m,100}^\kappa({\bb r})
\nonumber\\
&=&4\pi \beta^3 [Y_1^{m} ({\hat {  \bf p}})]^*  \left ( \frac{na_0}{2}\right )^3 \int x^2 dx
 j_{l}\left (\frac{pnax}{2(n+1)} \right ) e^{-{x}/{2}}
x L_{n-2-\kappa}^{3}\left (x\right ).~~~~~
\end{eqnarray}
Using the generating function of the Laguerre and Genegnbauer
polynomials, the Fourier transform can be carried out \cite{Pod29} ,
and we obtain
\begin{eqnarray}
{\tilde A}_{n1m,100}^\kappa({\bf p})&=&
4\pi [Y_l^{m} ({\hat {  \bf p}})]^*  \left ( \frac{na_0}{2}\right )^3 
\frac{{npa_0 (n+1)^{2}~ 2^6(n-\kappa) }}{((npa_0)^2+(n+1)^2)^3} 
\nonumber\\
& &
\times C_{n-2-\kappa}^{2} \left (\frac{(npa_0)^2-(n+1)^2}{(npa_0)^2+(n+1)^2}\right ).
\end{eqnarray}
Utilizing this result, we can write down the Fourier transform of the
transition density in the form
\begin{eqnarray}
{\tilde \rho}_{n1m,100} ({\bf p})
={\tilde R}_{n1m,100} (p)[Y_l^{m} ({\hat {  \bf p}})]^*,
\end{eqnarray}
where ${\tilde R}_{n1m,100} (p)$ is given by Eq.\ (\ref{rr}).  The transition matrix element becomes,
\begin{eqnarray}
\langle A_\lambda B_\lambda |V_{jk}|00\rangle({\bf r})&=&s \int \frac{d\bbox{p}}{(2\pi)^3} 
e^{i \bbox{p}\cdot \bbox{r}}
{\tilde R}_{n 1 m,100}^A(f_A(jk)p)  Y_{1 m}^* (\hat {\bf p})
\nonumber\\
& & 
\times ~~
{\tilde R}_{n'1-m,100}^B(f_B(jk)p)  Y_{1-m}^* (\hat {\bf p})
{\tilde v}_{jk}(p),
\nonumber\\
&=&
s\int \frac{p^2 dp}{(2\pi)^3} 
{\tilde R}_{n 1 m,100}^A(f_A(jk)p)  
{\tilde R}_{n'1-m,100}^B(f_B(jk)p) 
{\tilde v}_{jk}(p)
\nonumber\\
& &\times
\int\limits_{0}^{2\pi} d\phi \int\limits_{-1}^{1} d\mu
~e^{i \bbox{p}\cdot \bbox{r}}
Y_{1 m}^* (\hat {\bf p})  Y_{1 -m}^* (\hat {\bf p}).
\end{eqnarray}
The angular integral can be carried out and we obtain
\begin{eqnarray}
A(p,r)\equiv \int\limits_{0}^{2\pi} d\phi \int\limits_{-1}^{1} d\eta ~e^{i
\bbox{p}\cdot \bbox{r}}~Y_{1m}(\theta_p)~ Y_{1m}^*(\theta_p)
=\begin{cases}
 j_0(pr) -2j_2(pr)& ~~~({\rm for~} m=0), \cr
-[j_0(pr)+ j_2(pr)] & ~~~({\rm for~} m=1), \cr
 \end{cases}
\end{eqnarray}
where $j_{l}(x)$ are  spherical Bessel
function.  This gives Eq.\ (\ref{mat}) in Section VI.A.

\vspace {0.5cm}
\centerline{\bf Acknowledgment}
\vskip .5cm One of the authors (CYW) acknowledges the benefits of
tutorials at Princeton University from the late Professor
J. A. Wheeler whose ``polyelectrons" and ``nanosecond matter" had
either consciously or unconsciously stimulated the present
investigation.  For this reason the present article was written to
commemorate the Centennial Birthday of Professor J. A. Wheeler
(1911-2008).  The research was sponsored by the Office of Nuclear
Physics, U.S. Department of Energy.


\begin{thebibliography}{99}

\bibitem{Whe46}
J. A. Wheeler, ``Polyelectrons", Ann. New York Acad Sciences, 48, 219 (1946).

\bibitem{Hyl47}
E. A. Hylleraas, Phys. Rev. {\bf 71}, 491 (1947);
A. Ore, Phys. Rev. {\bf 71}, 913 (1947);

\bibitem{Mor73} D. L. Morgan and V. W. Hughes, Phys. Rev. {\bf A7},
1811 (1973).

\bibitem{Ho83}
Y. K. Ho, J. Phys. B {\bf 16}, 1503 (1983).

\bibitem{Gri89} J. J. Griffin, J. Phys. Soc. Jpn. {\bf 58},
S427(1989); J. J. Griffin, Phys. Rev. C{\bf 47}, 351
(1993); J. J. Griffin, Acta Phys. Polon. {\bf B27}  2087 (1996). 

\bibitem{Nus93}
S. Nussinov, Phys. Lett. B{\bf 314}, 397 (1993).

\bibitem{Koz93} P. M. Kozlowski and L. Adamowicz,
Phys. Rev. A {\bf 48}, 1903 (1993).

\bibitem{Zyg04}
B. Zygelman, A. Saenz, P. Froelich, and S. Jonsell,
Phys. Rev. A {\bf 69}, 042715 (2004).

\bibitem{Kolos}
W.~Kolos {\it et al}., Phys.~Rev.~A {\bf 11}, 1792 (1975).

\bibitem{Armour}
E.~A.~G.~Armour, J.~M.~Carr, and V.~Zeman, J. Phys. B {\bf 31}
L679 (1998); E.~A.~G.~Armour, and V.~Zeman, Int. J. Quantum Chem.
{\bf 74} 645 (1999).

\bibitem{Froelich}
P. Froelich, S. Jonsell, A. Saenz, B. Zylgelman, and A. Dalgarno,
Phys. Rev. Lett. {\bf 84}, 4577 (2000).

\bibitem{Jon01}
 S. Jonsell, A. Saenz, P. Froelich, B. Zylgelman, and A. Dalgarno,
Phys. Rev. A {\bf 64}, 052712 (2001).

\bibitem{Strasburger}
K. Strasburger, J. Phys. B {\bf 35}, L435 (2002).

\bibitem{Labzowsky}
L. Labzowsky, V. Sharipov, A. Prozorov, G. Plunien, and G. Soff,
Phys. Rev. A {\bf 72}, 022513, (2005).

\bibitem{Sharipov}
V. Sharipov {\it et al}., Phys. Rev. A {\bf 73}, 052503 (2006);
Phys. Rev. Lett {\bf 97}, 103005 (2006).

\bibitem{Cha01} M. Charlton and J. W. Humberston, {\it Positron
  Physics}, Cambridge University Press, Cambridge, (2001).

\bibitem{Whe88} J. A. Wheeler, ``Nanosecond Matter'' in {\it Energy in
  Physics, War, and Peace}, Kluwer Academic Publishers, 1988, page
  101.

\bibitem{Cas07}
D. B. Cassidy and A. P. Mills, Nature {\bf 449}, 195 (2007).
 
\bibitem{Amoretti}
M.~Amoretti {\it et al}., Nature, {\bf 419}, 456 (2002); Phys.
Lett. B, {\bf 578}, 23 (2004).

\bibitem{Gabrielse}
G.~Gabrielse {\it et al}., Phys. Rev. Lett., {\bf 89}, 213401,
222401 (2002); Adv. Atom Mol. Opt. Phys., {\bf 50}, 155 (2005).


\bibitem{And10}
G. B. Andresen $et~al.$,     Nature  {\bf   468},    673    (2010).


\bibitem{Won04}
C.~Y.~Wong, Phys. Rev. C {\bf 69}, 055202 (2004).

\bibitem{Lee08}
T. G. Lee, C. Y. Wong, and L.S. Wang, Chinese Physics {\bf 17}, 2897 (2008).

\bibitem{Whe37}
J. A. Wheeler, Phys. Rev. {\bf 52}, 1083 (1937).

\bibitem{Alm60} E. Almqvist, D. A. Bromley, and J. A. Kuehner,
Phys. Rev. Lett. {\bf 4}, 515 (1960); Y. Kondo, Y. Abe, and
T. Matsuse, Phys. Rev. {\bf C19}, 1356 (1979); G. R. Satchler, {\sl
Direct Nuclear Reactions}, (Oxford University Press, Oxford, 1983);
G. R. Satchler and W. G. Love, Phys. Rep. {\bf 55}, 183 (1979).

\bibitem{Tor03}
N. A. T\" ornqvist, Phys. Rev. Lett. {\bf 67}, 556 (1992); 
N. A. T\" ornqvist, Z. Phys. {\bf C61}, 525 (1994);
N. A. T\" ornqvist, Phys. Rev. Lett. 67, 556 (1991); 
N. A. T\" ornqvist, Phys. Lett. B 590, 209 (2004).

\bibitem{Bra04} E. Braaten and M. Kusunoki, Phys. Rev. D 69, 074005
(2004); E. Braaten and H.-W. Hammer, Phys. Rep. {\bf 428}, 259
(2006); E. Braaten and M. Lu, Phys. Rev. D 76, 094028 (2007);
E. Braaten and J. Stapleton, Phys. Rev. D 81, 014019
(2010).

\bibitem{Clo04}
F. E. Close and P. R. Page, Phys. Lett. B 578, 119 (2004).

\bibitem{Swa04} E. S. Swanson, Phys. Lett. B 588, 189 (2004).

\bibitem{Cho03}
S. K. Choi et al. (Belle Collaboration), Phys. Rev. Lett. 91,
262001 (2003).

\bibitem{Aco04}
D. Acosta et al. (CDF II Collaboration), Phys. Rev. Lett.
93, 072001 (2004);
V. M. Abazov et al. (D0 Collaboration), Phys. Rev. Lett.
93, 162002 (2004);
B. Aubert et al. (BABAR Collaboration), Phys. Rev. D 71,
071103 (2005);
B. Aubert et al. (BABAR Collaboration), Phys. Rev. D 77,
111101 (2008).


\bibitem{Hil53}
D. L. Hill and J. A. Wheeler, Phys. Rev. {\bf 89}, 1102 (1953).

\bibitem{Lan58}
L. D. Landau and E. M. Lifshitz, {\it Quantum Mechanics}, 
Pergamon Press, 1958, Eq.\ (38.7).


\bibitem{Sat83} G. R. Satchler and W. G. Love, Phys. Rep.  {\bf 55},
183 (1979).; F. Petrovich, Nucl. Phys. {\bf A251}, 143 (1975).

\bibitem{Bet57} 
H. A. Bethe and E. Salpeter, {\it Quantum Mechanics of
  one- and two-electron atoms}, Springer Verlag, Berlin, 1957.

\bibitem{Abr70}
M. Abramowitz and I. A. Stegun, {\it Handbook of Mathematical
Functions} (Dover Publications, Inc, New York) p785,
eq.~(22.12.7).

\bibitem{Bar81}
A. R. Barnett, 
Comp. Phys. Comm.  {\bf 27},  147   (1982).


\bibitem{Eis28}
R.~Eisenschitz and F.~London, Zeitschrift f\"ur Physik. {\bf 50},
24 (1928); F.~London, Nature, {\bf 17}, 516 (1929).

\bibitem{Cra06}
H. W.  Crater,  C.Y.  Wong, and P. Van Alstine,
Phys. Rev.  D{\bf 74},   054028 (2006).


\bibitem{Ber82}
V. B. Berestetskii, E. M. Lifshitz, and L. P. Pitaevskii, 
{\it Quantum Electrodynamics}, Pergamon Press, 1982.


\bibitem{Des63}
A. de-Shalit and I. Talmi,
{\it Nuclear Shell Theory}, Academic Press, N.Y. 1963.

\bibitem{Cha95}
L. Chatterjee and C. Y. Wong,
Phys. Rev. C{\bf 51}, 2125 (1995).


\bibitem{Won97}
C. Y. Wong and L. Chatterjee, Z. Phys. C {\bf 75}, 523 (1997).

\bibitem{PDG10}
K. Nakamura $et~al.$, Jour. Phys. G {\bf 37} 1 (2010).

\bibitem{Won94} C. Y. Wong, {\it Introduction to High-Energy Heavy-Ion
  Collisions}, World Scientific Publisher, 1994.

\bibitem{Won94a} 
See for example, Eq. (12.27) of Ref. \cite{Won94} for
  the inelastic cross section in the collision of two composite
  objects with constituents.

\bibitem{Pod29}
B. Podolsky and L. Pauling, Phys. Rev. 34, 109 (1929).

\end{thebibliography}
\end{document}